\documentclass[twocolumn]{aastex631}

\usepackage{amsmath}

\begin{document}

\title{Filter Design for Estimating the Stellar Metallicity of Metal-poor Stars from Gaia XP Spectra}

\correspondingauthor{Yang Huang}
\email{huangyang@ucas.ac.cn}

\author[0009-0000-6932-7894]{Ruifeng Shi}
\affiliation{School of Astronomy and Space Science, University of Chinese Academy of Sciences, Beijing 100049, China}

\author[0000-0003-3250-2876]{Yang Huang}
\affiliation{School of Astronomy and Space Science, University of Chinese Academy of Sciences, Beijing 100049, China}
\affiliation{National Astronomical Observatories, Chinese Academy of Sciences, Beijing 100101, China}

\author[0000-0001-8424-1079]{Kai Xiao}
\affiliation{School of Astronomy and Space Science, University of Chinese Academy of Sciences, Beijing 100049, China}

\author[0009-0006-7556-8401]{Chuanjie Zheng}
\affiliation{School of Astronomy and Space Science, University of Chinese Academy of Sciences, Beijing 100049, China}

\author[0009-0008-7479-0742]{Bowen Zhang}
\affiliation{School of Astronomy and Space Science, University of Chinese Academy of Sciences, Beijing 100049, China}
\affiliation{National Astronomical Observatories, Chinese Academy of Sciences, Beijing 100101, China}

\author[0009-0007-5610-6495]{Hongrui Gu}
\affiliation{School of Astronomy and Space Science, University of Chinese Academy of Sciences, Beijing 100049, China}
\affiliation{National Astronomical Observatories, Chinese Academy of Sciences, Beijing 100101, China}

\author[0000-0003-2086-0684]{Xinyi Li}
\affiliation{College of Physics and Electronic Engineering, Qilu Normal University, Jinan 250200, China}
\affiliation{Shandong Key Laboratory of Space Environment and Exploration Technology, China}

\author[0009-0008-2988-2680]{Huiling Chen}
\affiliation{Department of Astronomy, School of Physics, Peking University, Beijing 100871, China}
\affiliation{Kavli Institute for Astronomy and Astrophysics, Peking University, Beijing 100871, China}

%% Note that the \and command from previous versions of AASTeX is now
%% depreciated in this version as it is no longer necessary. AASTeX 
%% automatically takes care of all commas and "and"s between authors names.

%% AASTeX 6.31 has the new \collaboration and \nocollaboration commands to
%% provide the collaboration status of a group of authors. These commands 
%% can be used either before or after the list of corresponding authors. The
%% argument for \collaboration is the collaboration identifier. Authors are
%% encouraged to surround collaboration identifiers with ()s. The 
%% \nocollaboration command takes no argument and exists to indicate that
%% the nearby authors are not part of surrounding collaborations.

%% Mark off the abstract in the ``abstract'' environment. 
\begin{abstract}
The estimation of stellar atmospheric parameters for large-scale samples, particularly metal-poor stars, is a cornerstone of Galactic archaeology. In this work, we optimized a photometric filter design tailored to measuring stellar metallicities for very metal-poor stars with [Fe/H]\,$< -1$.The optimal configurations consist of a central wavelength $\lambda_{\rm c} = 3960$\,\AA\ with a bandwidth $\Delta\lambda = 80$\,\AA\ for giant stars, and $\lambda_{\rm c} = 3920$\,\AA\ with $\Delta\lambda = 80$\,\AA\ for dwarf stars. By applying these optimized filters to synthetic photometry derived from Gaia XP spectra, we inferred metallicities for both populations. Both internal and external validations demonstrate high precision across a wide metallicity range: 0.18--0.19\,dex for $-2 \le \rm [Fe/H] \le -1$, 0.23--0.33\,dex for $-3 \le \rm [Fe/H] \le -2$, and approximately 0.39\,dex for the most metal-poor regime, successfully extending down to $\rm [Fe/H] \approx -4$ for giant stars, $\rm [Fe/H] \approx -3.3$ for dwarf stars. Finally, we present a catalog of approximately 14.5 million metal-poor stars with robust $\rm [Fe/H]$ measurements, along with more than ten thousand red giant ultra metal-poor candidates with $\rm [Fe/H] < -4.0$, providing a valuable resource for exploring the early formation and chemical evolution of the Milky Way.

\end{abstract}

%% Keywords should appear after the \end{abstract} command. 
%% The AAS Journals now uses Unified Astronomy Thesaurus concepts:
%% https://astrothesaurus.org
%% You will be asked to selected these concepts during the submission process
%% but this old "keyword" functionality is maintained in case authors want
%% to include these concepts in their preprints.

%% From the front matter, we move on to the body of the paper.
%% Sections are demarcated by \section and \subsection, respectively.
%% Observe the use of the LaTeX \label
%% command after the \subsection to give a symbolic KEY to the
%% subsection for cross-referencing in a \ref command.
%% You can use LaTeX's \ref and \label commands to keep track of
%% cross-references to sections, equations, tables, and figures.
%% That way, if you change the order of any elements, LaTeX will
%% automatically renumber them.
%%
%% We recommend that authors also use the natbib \citep
%% and \citet commands to identify citations.  The citations are
%% tied to the reference list via symbolic KEYs. The KEY corresponds
%% to the KEY in the \bibitem in the reference list below. 

\section{Introduction} \label{sec:intro}
Understanding the formation and early evolution of the Milky Way is a central goal of Galactic archaeology (\citealt{2016ARA&A..54..529B}, \citealt{2018MNRAS.481.4093S}). This field relies on large stellar samples with well-determined kinematic and chemical properties to reconstruct the assembly history of the Galaxy. Although spectroscopic surveys can provide the most accurate measurements of stellar metallicity, their sample sizes remain limited compared to the vast stellar populations revealed by modern photometric and astrometric surveys, as represented by Gaia (\citealt{2016A&A...595A...1G},\citealt{2016A&A...595A...2G}).

The advent of Gaia has transformed this landscape. Gaia DR3 delivers low-resolution XP spectra \citep{2023A&A...674A..33G} for more than 200 million stars, covering a broad wavelength range from the near-UV to the near-infrared, i.e. 330 to 1050 nm. This unprecedented data set opens a new window for estimating stellar metallicities at a scale previously inaccessible. Recent studies have demonstrated that metallicity information can be extracted from Gaia XP spectra through two broad methodological approaches:
(1) data-driven or machine-learning techniques that exploit the full spectral information in combination with training labels \citep[e.g.,][]{2023ApJS..267....8A, 2023MNRAS.524.1855Z, 2024ApJ...963...95C, 2025ApJS..279....7Y}, and
(2) physics-motivated approaches that focus on specific spectral regions where metallicity-sensitive features, such as line blanketing in the blue, are expected to dominate \citep[e.g.,][]{2023A&A...674A.194B, 2024A&A...692A.115M, 2025A&A...700A..74O}.

In principle, physically motivated methods offer a particularly attractive path: they rely on transparent physical mechanisms and require fewer assumptions than high-dimensional machine-learning models. However, such approaches face a fundamental trade-off. While shorter wavelengths are more sensitive to metallicity through line blanketing, they also suffer from lower signal-to-noise ratios in Gaia XP spectra. An optimal metallicity indicator must therefore balance sensitivity against observational uncertainty.

This balance was first systematically explored by \citet{2024ApJ...968L..24X}, who designed optimized synthetic filters (with $\lambda_{\rm c} = 3790$\,\AA\, and $\Delta\lambda = 640$\,\AA) for metallicity estimation using Gaia XP spectra. However, their filter was specifically developed for dwarf stars, and the calibration sample is dominated by metal-rich stars, with limited coverage at [Fe/H] $< -1$ and extending down to $\sim -2.0$. Nevertheless, the most chemically informative stellar populations for probing the early Milky Way, metal-poor (MP; [Fe/H]~$<-1.0$), very metal-poor (VMP; [Fe/H]~$<-2.0$), and extremely metal-poor (EMP; [Fe/H]~$<-3.0$) stars, remain poorly sampled. Despite extensive efforts, the number of known EMP stars is still limited to only about $10^3$ objects, severely constraining studies of the Galaxy’s earliest evolutionary phases.

\begin{figure}
\centering
\plotone{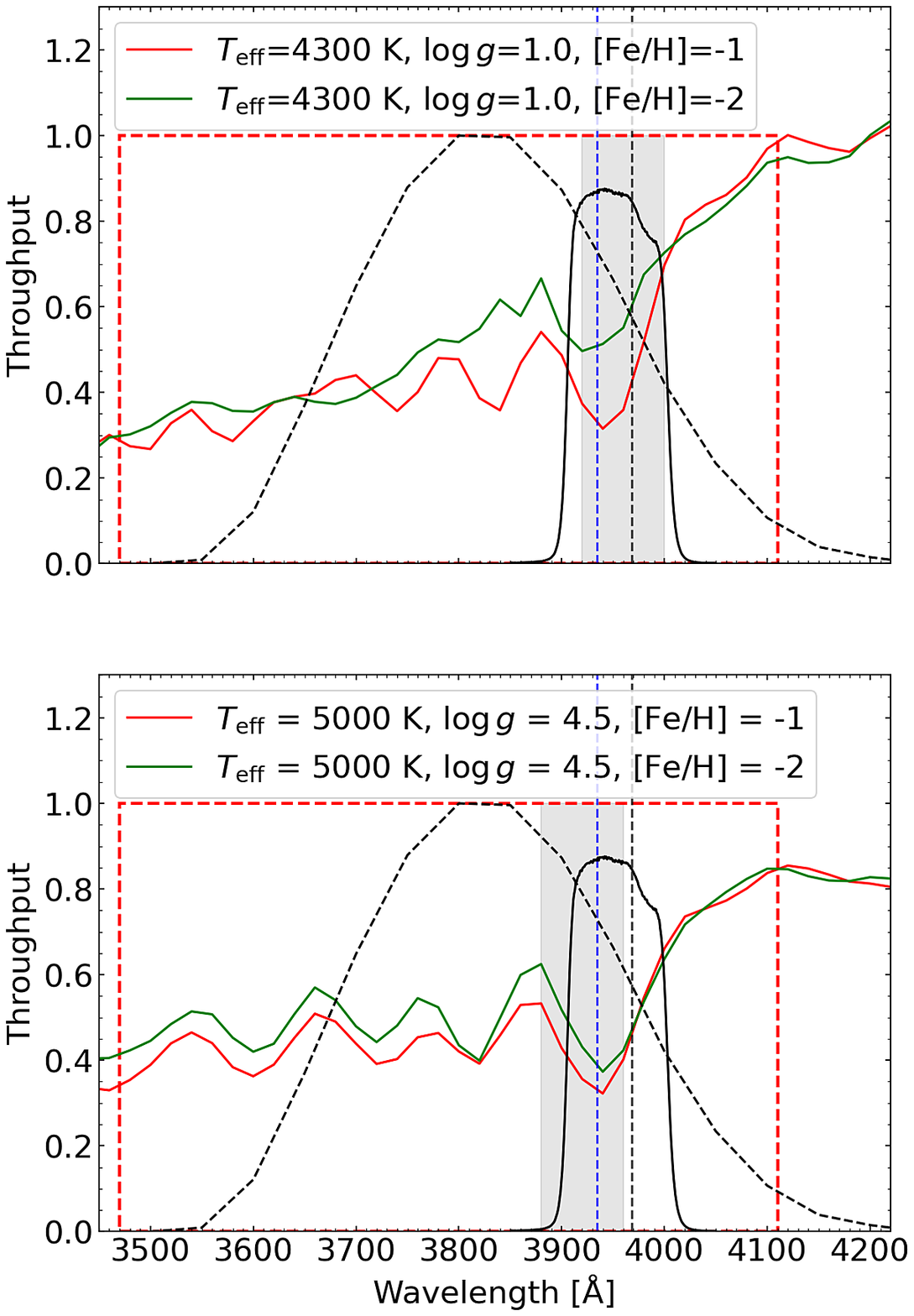}
\caption{Top panel: The gray shaded rectangle marks the optimal filter design for giant stars, derived using the algorithm described in Section 2. Two Gaia XP spectra are shown, with their atmospheric parameters from PASTEL/SAGA catalog labeled. The metallicity of the two spectra predicted by our method is $-$1.3, $-$2.2, respectively. The Ca H and Ca K lines are indicated by the black and blue vertical lines, respectively. The black solid line and black dashed line represent the transmission curves of the Pristine Ca H\&K band and SkyMapper $v$ band, respectively. The red rectangle shows the filter design from \citet{2024ApJ...968L..24X}. 
Bottom panel: Same as the top panel, but for the optimal filter design of dwarf stars. The metallicity of the two spectra from our work is $-$1.0, $-$1.8, respectively.
\label{fig1}}
\end{figure}

In this work, we specifically target this gap by optimizing filter design and metallicity estimation for stars with [Fe/H] $< -1$, separately for dwarf and giant stars, extending the metal-poor limit for dwarfs to $\mathrm{[Fe/H]} \approx -3.5$ and enabling dedicated optimization for giants down to $\mathrm{[Fe/H]} \approx -4.0$. By re-evaluating the sensitivity–noise balance in Gaia XP spectra (\citealt{2021A&A...652A..86C}, \citealt{2024ApJS..271...13H}) for metal-poor stars, we identify filter configurations that maximize metallicity precision while maintaining robustness against observational uncertainties, as shown in Figure~\ref{fig1}. Our approach is grounded in the physical origin of metallicity signatures and avoids unnecessary model complexity.

The resulting methodology enables significantly improved metallicity estimates for MP stars and provides an efficient pathway to identifying large samples of VMP and EMP candidates. Applying this framework to Gaia XP data is expected to yield metallicities for tens of thousands of EMP candidates, substantially enlarging the available sample for studies of the early formation and chemical evolution of the Milky Way.

This paper is organized as follows. Section 2 describes the construction of the training sample from the PASTEL and SAGA catalogs and the metallicity estimation method. Section 3 presents validation tests of the derived metallicities. Section 4 discusses the achievable precision and limitations of the method, and Section 5 summarizes our main conclusions.

\section{Data and method} \label{sec:method}

The primary goal of this work is to identify the optimal filter configuration for precise stellar metallicity estimation. To do so, we first describe the derivation of synthetic colors from Gaia XP spectra and the correction for interstellar reddening in Section 2.1. In Section 2.2, we analyze the sensitivity and uncertainty of metallicity estimates using a high-quality reference sample as the training set, construct metallicity-dependent stellar loci, and derive stellar metallicities using a maximum-likelihood approach. Finally, in Section 2.3, we apply the above procedure across the $\lambda_{\rm c}$--$\Delta\lambda$ parameter space to determine the optimal filter configuration. The following subsections present each of these steps in detail.

\subsection{Synthetic colors}
Deriving synthetic stellar colors is the first step toward estimating photometric metallicity. These colors are obtained by convolving filter transmission curves with observed spectra, for which we use Gaia XP data. In this work, we compute the synthetic colors $G_{\rm BP}-G_{\rm RP}$ and $m-G_{\rm BP}$, where $m$ denotes the magnitude in the optimized filter design described below. The color $G_{\rm BP}-G_{\rm RP}$ traces effective temperature and is useful for selecting F/G/K stars for metallicity estimation, while $m-G_{\rm BP}$ is primarily sensitive to [Fe/H]. 

For the Gaia BP and RP bands, the Vega magnitude system is recommended by the Gaia Collaboration\footnote{\url{https://gaiaxpy.readthedocs.io/en/latest/_static/images/PhotometricSystem_table.pdf}}. 
The Vega magnitude is defined as
\begin{equation}\label{eq1}
m_{\rm VEGA} = -2.5 \log 
\frac{\int_{\lambda_{i}}^{\lambda_{f}} f(\lambda)\,S(\lambda)\,\lambda\, d\lambda}
{\int_{\lambda_{i}}^{\lambda_{f}} S(\lambda)\,\lambda\, d\lambda}
+ ZP ,
\end{equation}
where $\lambda_{i}$ and $\lambda_{f}$ denote the wavelength range of a given band, 
$f(\lambda)$ is the flux density from the Gaia XP spectrum, and $S(\lambda)$ is the filter transmission curve. 
The zero point ($ZP$) is determined from the Vega spectrum as
\begin{equation}\label{eq2}
ZP = +2.5 \log 
\frac{\int_{\lambda_{i}}^{\lambda_{f}} f_{\rm vega}(\lambda)\,S(\lambda)\,\lambda\, d\lambda}
{\int_{\lambda_{i}}^{\lambda_{f}} S(\lambda)\,\lambda\, d\lambda},
\end{equation}
where $f_{\rm vega}(\lambda)$ is the flux density from the CALSPEC Vega spectrum\footnote{\url{https://svo2.cab.inta-csic.es/theory/fps/morefiles/vega.dat}}.

By convolving the Gaia BP and RP transmission curves with the Gaia corrected XP spectra (\citealt{2024ApJS..271...13H}), we obtained the corresponding synthetic BP and RP magnitudes. The original Gaia XP spectra are affected by systematic errors that depend primarily on the $G_{\rm BP} - G_{\rm RP}$ color and the $G$ magnitude. \citet{2024ApJS..271...13H} established empirical relations linking these systematics to source properties, enabling effective corrections, particularly in the near-ultraviolet. Compared with the color $G_{\rm BP} - G_{\rm RP}$ from GaiaXPy (\citealt{daniela_ruz_mieres_2022_7015044}), the scatter is around 0.005 mag, which shows the roubustness of our method in calculating the synthetic magnitude and color. 

The synthetic magnitude of the designed filter $m$ is computed following the same procedure as done by \citet{2024ApJ...968L..24X}. For the trial filter (assumed to have a top-hat transmission profile), we convolved the transmission curve with the Gaia XP spectra and computed the synthetic magnitude using Equation~(\ref{eq1}).

During the computation of synthetic magnitudes, interstellar reddening must be corrected to obtain intrinsic stellar colors. 
The line-of-sight extinction is taken from a three-dimensional dust map, with $E(B-V)$ values adopted from \citet{2025ApJS..280...15W}. 
For the designed filter, whose bandwidth is relatively narrow, we assume an extinction law with $R_{V}=3.1$ \citep{2024zndo..13333814G} to compute the reddening coefficient. 
In contrast, the Gaia $G_{\rm BP}$ and $G_{\rm RP}$ bands are very broad, and their reddening coefficients depend on the spectral energy distribution (SED), particularly on $T_{\rm eff}$ as well as the amount of reddening. 
Therefore, we adopt the temperature- and reddening-dependent extinction coefficients calibrated by \citet{2023ApJS..264...14Z} to correct the reddening effect of $G_{\rm BP}-G_{\rm RP}$. The effective temperatures are adopted from the PASTEL/SAGA.

\subsection{Metallicity estimates}

Deriving photometric metallicities and assessing their uncertainties requires a training sample with accurate spectroscopic metallicities. We therefore compile a reference sample from the Stellar Abundances for Galactic Archaeology (SAGA) database \citep{2008PASJ...60.1159S} and the PASTEL catalog \citep{2022A&A...663A...4S}. Both PASTEL and SAGA are literature-based compilations that aggregate atmospheric parameters derived from high-resolution optical spectroscopy ($R > 25,000$) across multiple studies, and thus include measurements obtained under both local thermodynamic equilibrium (LTE) and non-LTE assumptions.
The combined sample contains 24,160 stars with reported atmospheric parameters, with mean values adopted for sources with multiple literature measurements. As these catalogs are heterogeneous, systematic differences between measurements from different studies are expected. Previous work \citep{2022ApJ...925..164H} has quantified these differences and shown that, although the scatter increases toward lower metallicity, it remains within $\sim$0.10 dex even at [Fe/H] $\approx -3.5$.
In principle, a self-consistent training set based on homogeneous non-LTE analyses (e.g., GALAH) would be desirable. However, such surveys currently lack sufficient coverage in the extremely metal-poor regime ([Fe/H] $< -3$). We therefore adopt the combined PASTEL+SAGA sample to ensure adequate coverage at the metal-poor end. As shown in Figure A1, the metallicities of common stars between high-resolution literature samples and GALAH are in good agreement, with only minor offsets at low metallicity that can be readily calibrated. A more detailed discussion is presented in Section 4.

\begin{figure*}
\centering
\plotone{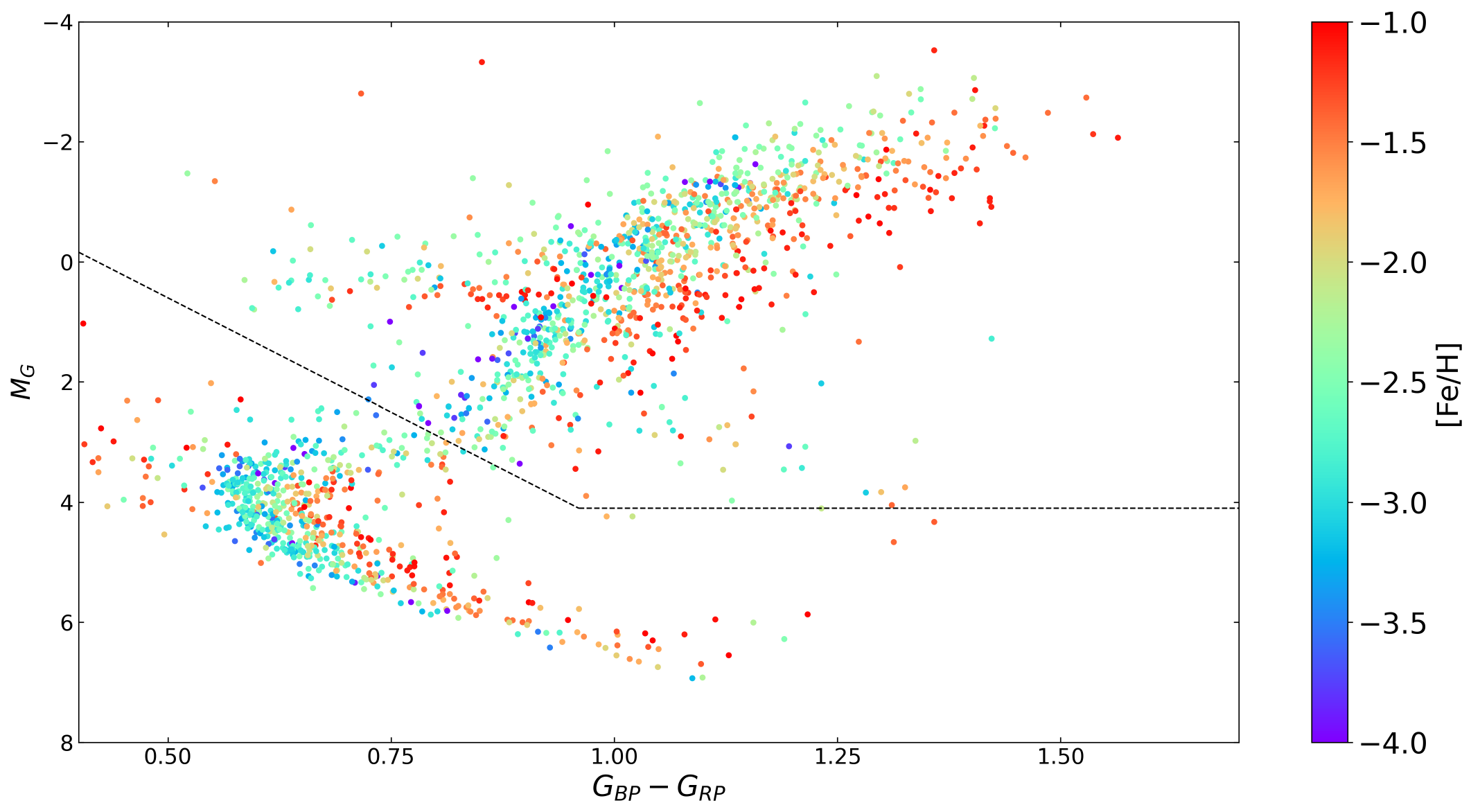}
\caption{The distribution of the training sample in the $M_{\rm G}$ versus $(G_{\rm BP}-G_{\rm RP})$ plane, color coded by the metallicity from SAGA Database/PASTEL catalog. The dashed lines represent the cuts $M_{G}=-3.20+7.60(G_{\rm BP}-G_{\rm RP})$ or $M_{G}=4.1$ \citep{2022ApJ...925..164H}, used to separate the giant and dwarf stars, respectively.
\label{fig2}}
\end{figure*}

We then cross-match the combined PASTEL+SAGA sample with the corrected Gaia XP spectra from \citet{2024ApJS..271...13H}, resulting in a sample of 16,412 stars. Using the method described in Section 2.1, we calculate the dereddened colors $G_{\rm BP}-G_{\rm RP}$ and $m-G_{\rm BP}$ for all stars in this sample, where $m$ denotes a trial metallicity-sensitive filter.
The main goal of this work is to use the designed filter to estimate the metallicities of metal-poor stars; therefore, only stars with [Fe/H] $\leq -1$ are retained. 
We further define a \emph{golden sample} using the following criteria: 
(1) photometric uncertainties $\sigma_{\rm BP} < 0.01$ and $\sigma_{\rm RP} < 0.01$; 
(2) $0 < E(B-V) < 0.2$, to limit the impact of reddening corrections while retaining a sufficiently large training sample;
(3) $0.4 \leq G_{\rm BP}-G_{\rm RP} \leq 1.8$, corresponding to FGK-type stars, i.e., the temperature range where the filter is most sensitive to metallicity.
Finally, 2,212 metal-poor stars constitute the training sample, including 1,582 giants and 630 dwarfs.
Figure~\ref{fig2} shows the distribution of the training sample in the color--absolute magnitude diagram. The number of giant stars in different metallicity bins is as follows: 715 for $-2 \le \rm [Fe/H] \le -1$, 657 for $-3 \le \rm [Fe/H] \le -2$, 196 for $-4 \le \rm [Fe/H] \le -3$, and 14 for $\rm [Fe/H] < -4$. The corresponding numbers for dwarf stars are 228, 267, 129, and 6, respectively. Compared to \citet{2024ApJ...968L..24X}, our sample provides substantially improved coverage at the low-metallicity end, enabling more robust metallicity estimates for VMP stars.
Using the empirical boundaries defined in \citealt{2022ApJ...925..164H} in the $M_G$–$(G_{\rm BP}-G_{\rm RP})$ plane, we separate dwarf and giant stars and train them independently to derive stellar metallicities.

To estimate metallicity, the key step is to construct metallicity-dependent stellar loci based on synthetic colors. 
We define the metallicity sensitivity as the gradient of the color $(m-G_{\rm BP})$ with respect to [Fe/H], which serves as a key indicator of the color range over which metallicity-dependent stellar loci can be reliably constructed. The synthetic magnitude of the designed filter $m$, combined with $G_{\rm BP}$, exhibits strong sensitivity to [Fe/H] for metal-poor stars only within a limited color range. Outside this range, the sensitivity rapidly degrades and becomes insufficient for reliable metallicity estimation. Based on this behavior, evaluated across a series of trial filters $m$ (whose design will be described in the next subsection), we identify the effective $G_{\rm BP}-G_{\rm RP}$ ranges to be $0.9$--$1.8$ for giants and $0.65$--$1.2$ for dwarfs. This corresponds to 1024 giant stars and 168 dwarf stars in our sample. The number of giant stars in different metallicity bins is 461 for $-2 \le \rm [Fe/H] \le -1$, 422 for $-3 \le \rm [Fe/H] \le -2$, 133 for $-4 \le \rm [Fe/H] \le -3$, and 8 for $\rm [Fe/H] < -4$. The corresponding numbers for dwarf stars are 100, 48, and 20, with no stars at $\rm [Fe/H] < -4$.

\begin{figure*}
\centering
\plotone{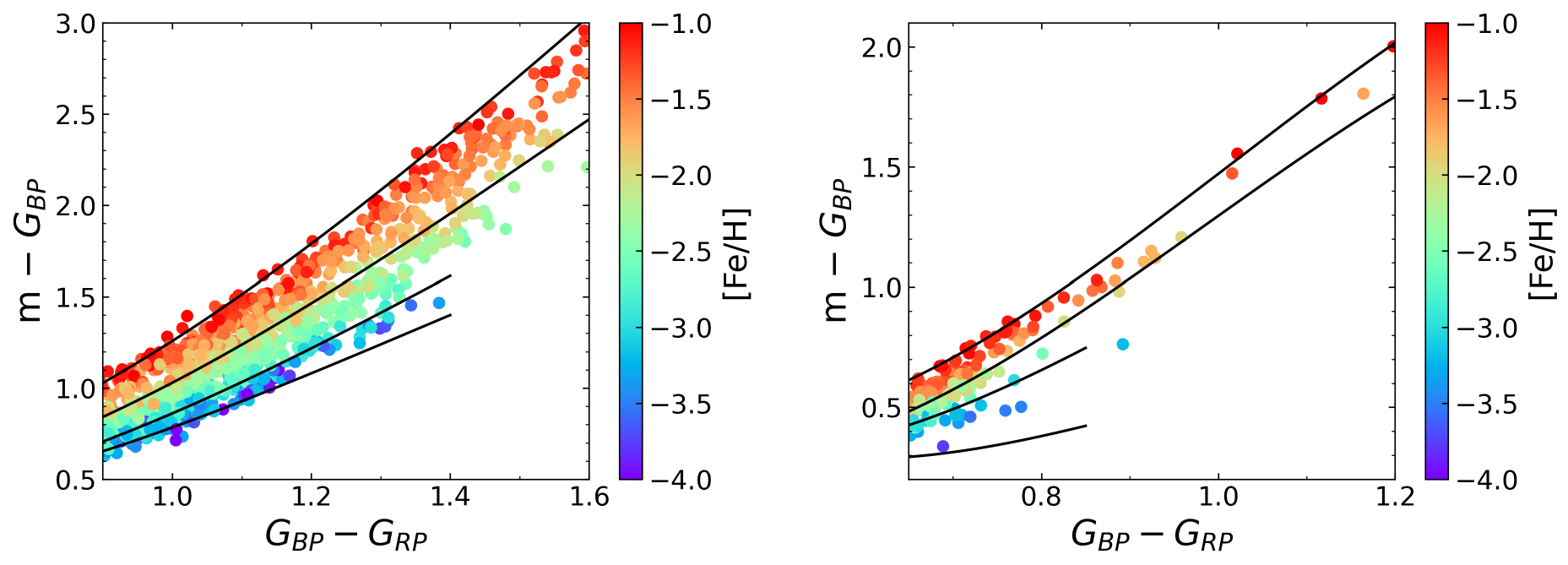}
\caption{The distribution of the metal-poor giant stars (left panel) and metal-poor dwarf stars (right panel) in the $m-G_{BP}$ versus $G_{\rm BP}-G_{\rm RP}$ plane, color coded by the metallicity. The black lines represent the best fitting [Fe/H] values utilizing Equation (3). From top to bottom, the [Fe/H] values of the four lines are $-$1, $-$2, $-$3 $-$4, respectively. 
\label{fig3}}
\end{figure*}

For a given filter magnitude $m$, we construct metallicity-dependent stellar loci. Figure~\ref{fig3} shows an example using the final optimized filter design, revealing a clear metallicity sequence in the $(m-G_{\rm BP})$ versus $(G_{\rm BP}-G_{\rm RP})$ plane. Following previous studies (\citealt{2015ApJ...799..134Y}; \citealt{2022ApJ...925..164H}; \citealt{2023ApJ...957...65H}; \citealt{2024RAA....24d5015S}), we adopt third-order two-dimensional polynomials to fit $(m-G_{\rm BP})$ as a function of $(G_{\rm BP}-G_{\rm RP})$ and [Fe/H], separately for these 168 dwarfs and 1024 giants:

\begin{equation}\label{eq3}
\begin{split}
(m-G_{\rm BP}) =\;& a_{0,0}+a_{0,1}y+a_{0,2}y^{2}+a_{0,3}y^{3}+a_{1,0}x \\
&+a_{1,1}xy+a_{1,2}xy^{2}+a_{2,0}x^{2}\\
&+a_{2,1}x^{2}y+a_{3,0}x^{3},
\end{split}
\end{equation}
where $x = (G_{\rm BP}-G_{\rm RP})$ and $y = \mathrm{[Fe/H]}$. 
The fitted coefficients are listed in Table~1. 
Based on these stellar loci, the metallicities of metal-poor dwarf and giant stars are derived using a maximum-likelihood approach. 
For a given star, the likelihood is defined as
\begin{equation}\label{eq4}
L_{c} = \frac{1}{\sqrt{2\pi}\,\sigma_{c_{1}}}
\exp\!\left[-\frac{(c_{1}-c_{2})^{2}}{2\sigma_{c_{1}}^{2}}\right],
\end{equation}
where $c_{1}=(m-G_{\rm BP})$ is the observed color derived from Gaia XP spectra and assumed to follow a Gaussian distribution, and $c_{2}$ is the color predicted by the metallicity-dependent stellar locus (Equation~\ref{eq3}). 
The metallicity grid spans $-4.0 \leq \mathrm{[Fe/H]} \leq -1.0$ with a step of 0.02\,dex when predicting $(m-G_{\rm BP})$. The best-fit [Fe/H] corresponds to the maximum likelihood, and the metallicity uncertainty (half of the 68\% interval) is estimated from the resulting probability distribution function (PDF).

\begin{table}
\begin{center}
\caption{ Fitting coefficients for giant and dwarf stars.}\label{Tab1}
\begin{tabular}{crr}
  \hline\noalign{\smallskip}
Coeff.    & Giant star    & Dwarf star \\
  \hline\noalign{\smallskip}
$a_{0,0}$    & 0.48311047    & 0.99959239   \\
$a_{0,1}$    & $-$0.21075645    & 0.35602951    \\
$a_{0,2}$    & $-$0.06799448    & 0.12978635    \\ 
$a_{0,3}$    & $-$0.00678829    & 0.01417430    \\
$a_{1,0}$    & $-$1.53752633    & $-$2.87334996    \\
$a_{1,1}$    & 0.26971967    & $-$0.11940625    \\
$a_{1,2}$    & 0.05374289    & $-$0.00269902     \\
$a_{2,0}$    & 3.10570299    & 5.16690551    \\
$a_{2,1}$    & 0.16785122    & 0.35415719     \\
$a_{3,0}$    & $-$0.56710603    & $-$1.38094273   \\
  \noalign{\smallskip}\hline
\end{tabular}
\end{center}
\end{table}

\subsection{The best filter design}

In order to identify the best filter configuration, we search in the range 
$3360\,\text{\AA} \le \lambda_{\rm c} \le 4140\,\text{\AA}$ 
with a step of 20\,\AA, and 
$40\,\text{\AA} \le \Delta\lambda \le (2\lambda_{\rm c}-3360)\,\text{\AA}$ 
with a step of 40\,\AA\, where $\lambda_{\rm c}$ and $\Delta\lambda$ denote the central wavelength and bandwidth of the filter, respectively. We test with such step because the resolution of Gaia XP spectra is $\Delta\lambda$ = 20\AA.
For each trial filter, the magnitude $m$ is derived, and thus the color 
$(m-G_{\rm BP})$ is constructed. 
Using a maximum likelihood approach, the metallicity of each star in the final sample is estimated from the colors $(m-G_{\rm BP})$ and $G_{\rm BP}-G_{\rm RP}$. 
By comparing with the [Fe/H] values from the original catalog, we derive the offset and scatter, and select the optimal filter as the one yielding the lowest scatter.

\begin{figure*}
\centering
\plotone{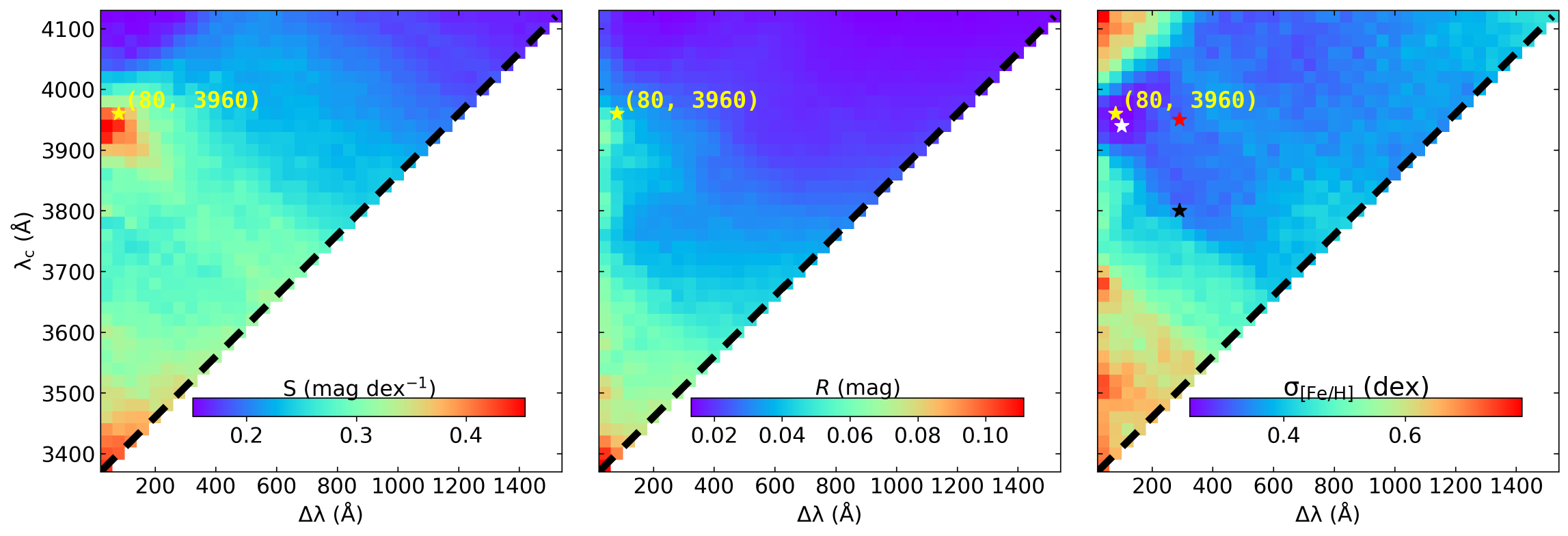}
\caption{Left panel: Metallicity sensitivity $S$ in the $\lambda_{\rm c}$--$\Delta\lambda$ plane for giant stars. 
The highest sensitivity appears in the Ca H\&K region around $\lambda_{\rm c}\!\sim\!3900$\,\AA\ and in the 
bluest region near $\lambda_{\rm c}\!\sim\!3400$\,\AA. 
Middle panel: Same as the left panel, but showing the scatter of the fitting residuals $R$. 
Right panel: Same as the left panel, but showing the precision of the estimated metallicity, 
$\sigma_{\rm [Fe/H]}$. The optimal filter parameters are marked by a yellow star. 
The red, black, and white stars denote the positions of the SAGES $v$ band, the SkyMapper $v$ band, 
and the Pristine Ca H\&K band, respectively.
\label{fig4}}
\end{figure*}

The results for giant stars are shown in Figure~\ref{fig4}. 
Two local sensitivity peaks can be clearly identified, with central wavelengths around 3400\,\AA\ and 3900\,\AA, respectively. 
The first peak arises because this wavelength range corresponds to the region where the overall metallicity-driven flux redistribution in XP spectra is strongest. 
The second peak is related to the Ca H\&K bands, which remain highly sensitive to metallicity even for metal-poor stars where most metal lines become weak. 
However, despite its strong sensitivity, the bluer peak suffers from significantly increased XP measurement uncertainties toward the blue edge of the spectrum. 
This leads to larger fitting residuals and ultimately degrades the precision of the inferred [Fe/H] (see middle and right panels of Fig.~\ref{fig4}). 
Therefore, the optimal filter design is located around the Ca H\&K region, with 
$\lambda_{\rm c}=3960$\,\AA\ and $\Delta\lambda=80$\,\AA.

In contrast to \citet{2024ApJ...968L..24X}, which focuses on metal-rich stars where many metal lines are strong, this work targets metal-poor stars with intrinsically weak spectral features. 
In this regime, only the strongest metal lines provide robust metallicity information. 
Our results indicate that focusing on the most prominent spectral features is essential for accurate photometric metallicity estimates of metal-poor stars.
Table~\ref{Tab2} lists the precision of metallicity derived from our optimal filter and from other custom filters, demonstrating that our design achieves the highest precision for giant stars. 
We note that the Ca H\&K filter adopted by the Pristine survey is already very close to our optimal solution. 
Our design is slightly narrower, which increases the sensitivity to metallicity and leads to a modest improvement in the final [Fe/H] precision.

\begin{figure*}
\centering
\plotone{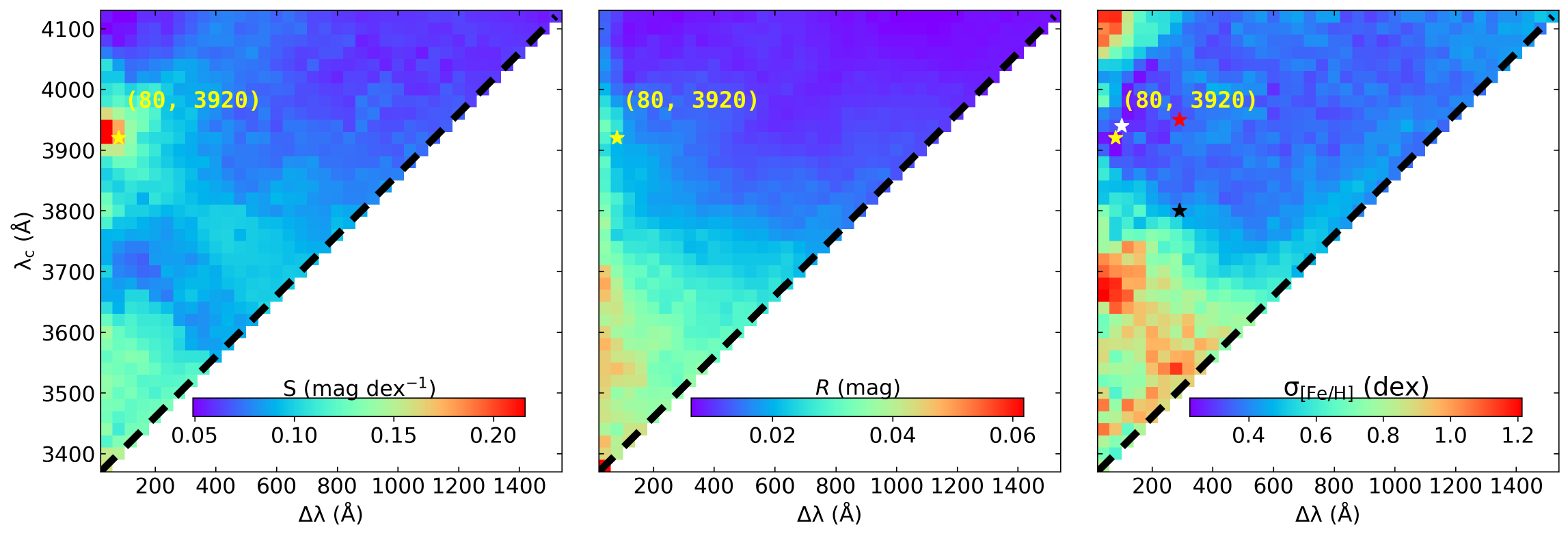}
\caption{Similar to Figure~\ref{fig4}, but for the dwarf stars. The best parameter of the designed filter is indicated by the yellow star, as shown in the three panels.
\label{fig5}}
\end{figure*}

For dwarf stars, the overall design is similar. 
The best filter parameters are 
$\lambda_{\rm c}=3920$\,\AA\ and $\Delta\lambda=80$\,\AA, as shown in Figure~\ref{fig5}. 
Similarly, the optimal central wavelength lies within the Ca H\&K region, consistent with expectations. Table~\ref{Tab3} lists the precision of metallicity derived from our optimal filter and from other custom filters for dwarf stars.

\section{Validations} \label{sec:result}

The main purpose of our work is to design the best filter in the ultraviolet band to provide the most precise [Fe/H] measurement for the metal-poor stars. In order to examine the precision of the metallicity derived using our approach, both internal and external validations are done.

\subsection{Internal validations}

\begin{figure*}
\centering
\plotone{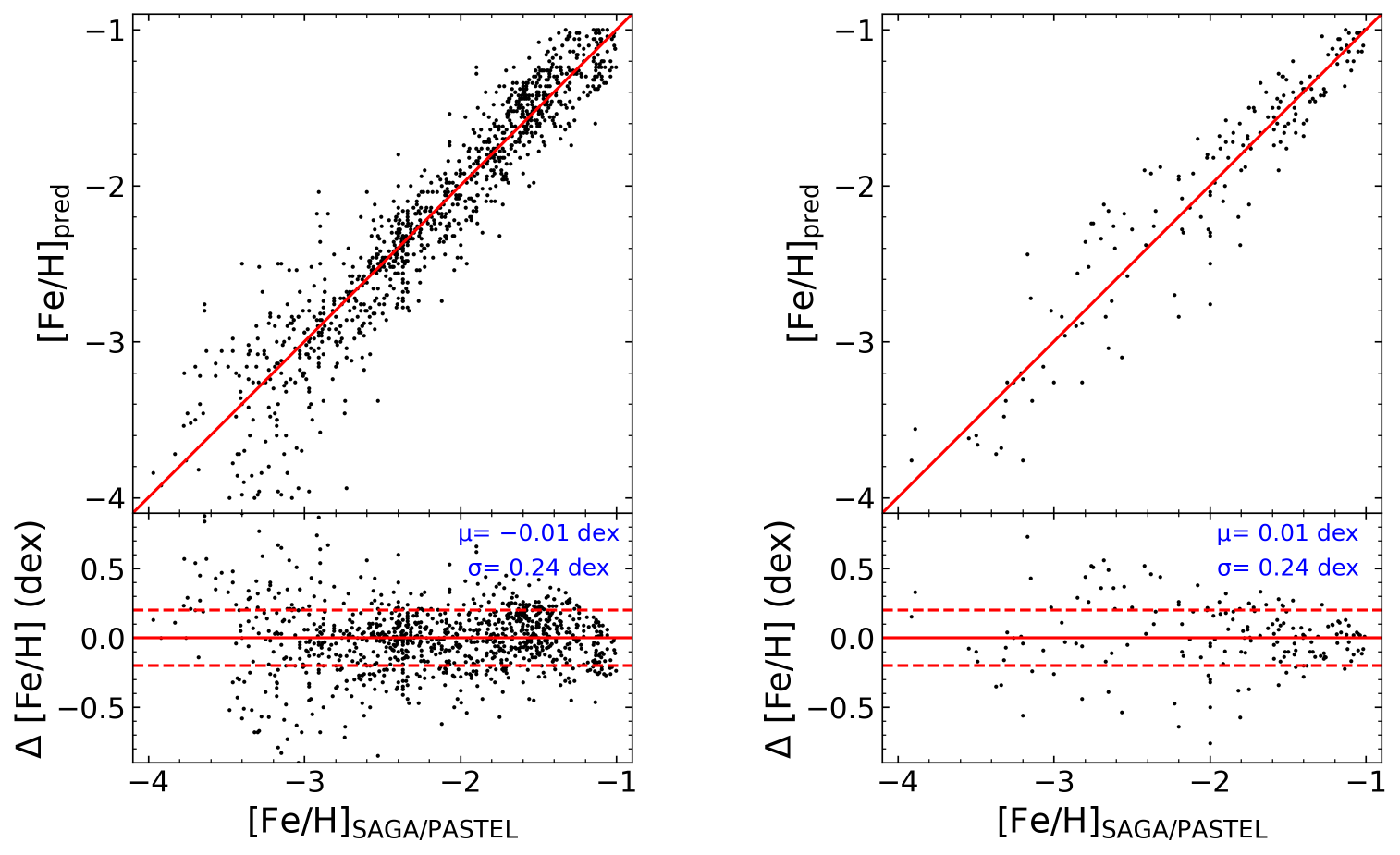}
\caption{The internal validation of the metallicity for giant (left panel) and dwarf (right panel)  stars. 
The lower part of each panel shows the metallicity difference (predicted metallicity minus that 
from the SAGA database or the PASTEL catalog) as a function of the catalog metallicity. 
Lines of $\Delta$[Fe/H] $= \pm 0.2$\,dex are plotted to guide the eye. 
The median and scatter of the metallicity differences are labeled in the upper-right corner of the lower panels.
\label{fig6}}
\end{figure*}

From the above analysis, the best-designed filters are adopted to estimate the metallicity of giant and dwarf stars, respectively. All 1024 giant stars and 168 dwarf stars used to define the color–metallicity relations are included in the internal validation. It shows good agreement between the predicted [Fe/H] and [Fe/H] from the original catalog, with a scatter of 0.24 dex for both giant and dwarf stars (as shown in Figure~\ref{fig6}). More details are listed in Table~\ref{Tab4}. The precision of the predicted metallcity is quite high, with typical values around 0.18---0.19 dex for stars with $-2 \le \rm[Fe/H] \le -1$, 0.23---0.33 dex for stars with $-3 \le \rm[Fe/H] \le -2$, around 0.39 dex for stars with $-4 \le \rm[Fe/H] \le -3$. This shows that our method of determining [Fe/H] can reach fairly high precision even for EMP stars. These stars span an apparent $G$-magnitude range of 4.8--16.8, broadly representative of the Gaia XP sample. For giant stars in the $-4 \le \mathrm{[Fe/H]} \le -3$ regime, the residuals show a slight negative systematic offset, indicating mild underestimation of their metallicities.

\begin{table}
\begin{center}
\caption{Metallicity sensitivity $S$, the scatter of fitting residuals $R$, and metallicity precision $\sigma_{\rm [Fe/H]}$ for giant stars, shown for the best filter design of this work and three custom filters.}\label{Tab2}
\begin{tabular}{cccc}
  \hline\noalign{\smallskip}
   Filter &  $S$    &   $R$  & $\sigma_{\rm [Fe/H]}$    \\
         &  (mag dex$^{-1}$)  &  (mag)  & (dex) \\
  \hline\noalign{\smallskip}
 This work  & 0.40    &  0.04    & 0.24 \\
 SAGES  &  0.29   & 0.03    & 0.34  \\ 
 SkyMapper  & 0.28   & 0.03    & 0.34    \\
 Pristine  & 0.39   & 0.04   &  0.27  \\

  \noalign{\smallskip}\hline
\end{tabular}
\end{center}
\end{table}

\begin{table}
\begin{center}
\caption{Similar to Table 2, but for the dwarf stars.}\label{Tab3}
\begin{tabular}{cccc}
  \hline\noalign{\smallskip}
   Filter &  $S$    &   $R$  & $\sigma_{\rm [Fe/H]}$    \\
         &  (mag dex$^{-1}$)  &  (mag)  & (dex) \\
  \hline\noalign{\smallskip}
 This work  & 0.18    &  0.02    & 0.24 \\
 SAGES  &  0.10   & 0.02    & 0.34  \\ 
 SkyMapper  & 0.09   & 0.02    & 0.42    \\
 Pristine  & 0.14   & 0.02   &  0.28  \\

  \noalign{\smallskip}\hline
\end{tabular}
\end{center}
\end{table}

\begin{table}
\begin{center}
\caption{ The precision of the metallicity for giant and dwarf stars from internal validations.}\label{Tab4}
\begin{tabular}{clccc}
  \hline\noalign{\smallskip}
    & $\sigma_{1}$    & $\sigma_{2}$ \\
        & (dex)   & (dex) \\
  \hline\noalign{\smallskip}
$-2 \le \rm[Fe/H] \le -1$    & 0.18   & 0.19   \\
$-3 \le \rm[Fe/H] \le -2$    & 0.23    & 0.33    \\
$-4 \le \rm[Fe/H] \le -3$    & 0.39    & 0.29$^{\rm a}$ \\ % 这里手动加个上标 a

  \noalign{\smallskip}\hline
\end{tabular}
\end{center}
% 在 tablecomments 里增加对 a 的解释
\tablecomments{Here $\sigma_{i}$ ($i$=1, 2) is the precision of the predicted metallicity for the giant and dwarf stars, respectively. The precision of the predicted metallicity in different [Fe/H] bins is listed in each row. \\
$^{\rm a}$ This value might be underestimated due to the limited number of calibration stars available in this specific metallicity bin.}
\end{table}

\subsection{External validations}

To evaluate the robustness of our photometric metallicity estimates, we perform extensive external validations for giant and dwarf stars separately.

\begin{figure*}
\centering
\plotone{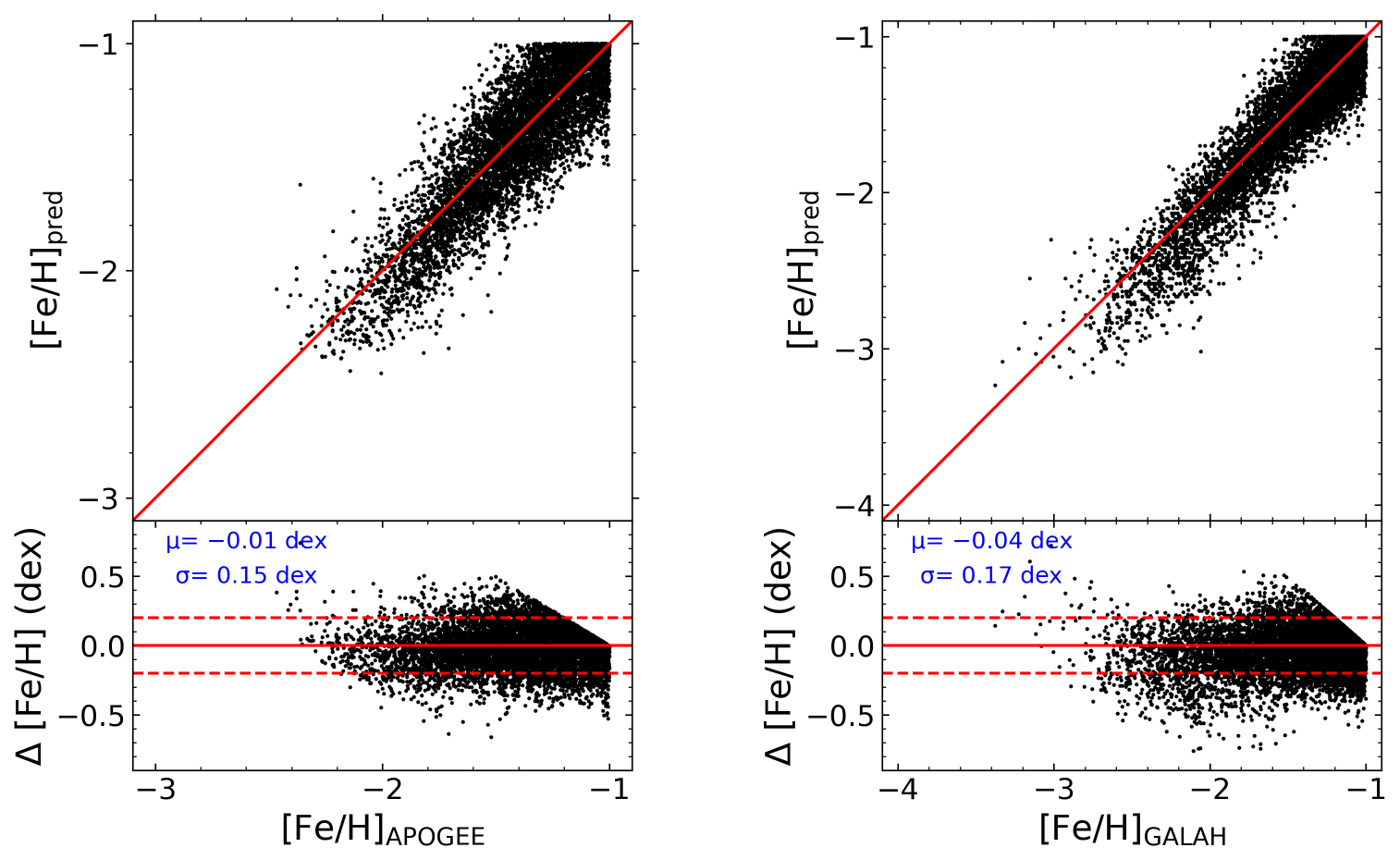}
\caption{Similar to Figure~\ref{fig6} but for the external validation of the metallicity for the giant stars using APOGEE DR17 (left panel) and GALAH DR4 (right panel).
The “arrow”-like feature in both panels arises from the upper limit of our method ($\mathrm{[Fe/H]} = -1$) combined with the selection cut of the comparison sample ($\mathrm{[Fe/H]} < -1$), which together produce this characteristic pattern.
\label{fig7}} 
\end{figure*}

\begin{figure}
\centering
\plotone{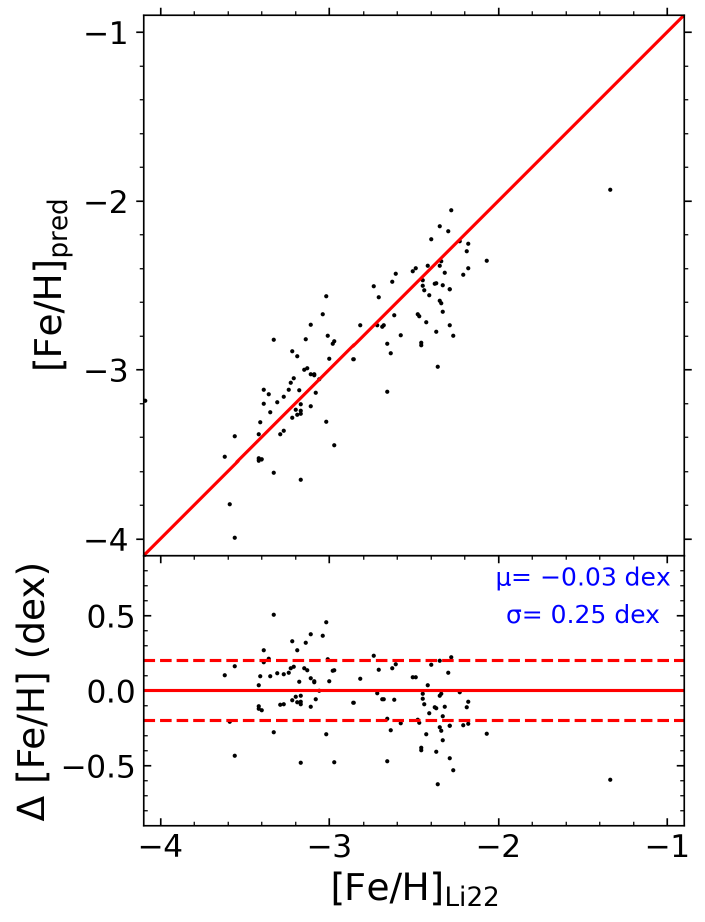}
\caption{Similar to Figure~\ref{fig6} but for the external validation of the metallicity for the giant stars using the catalog of \citet{2022ApJ...931..147L}.
\label{fig8}}
\end{figure}

\begin{figure}
\centering
\plotone{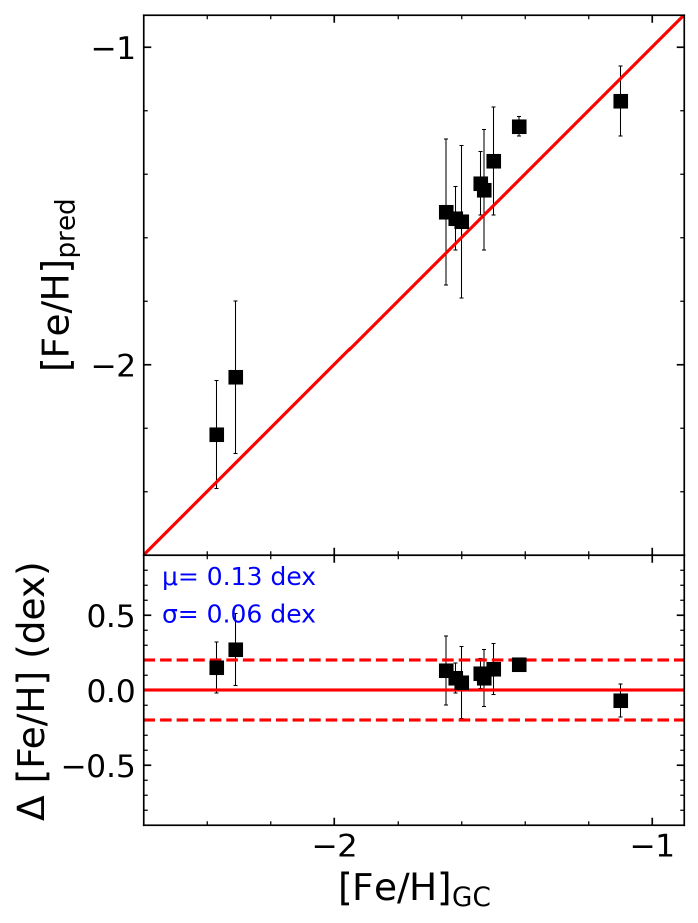}
\caption{Similar to Figure~\ref{fig6} but for the external validation of the metallicity for the giant stars using the member stars of the globular cluster.
\label{fig9}}
\end{figure}

\begin{figure*}
\centering
\plotone{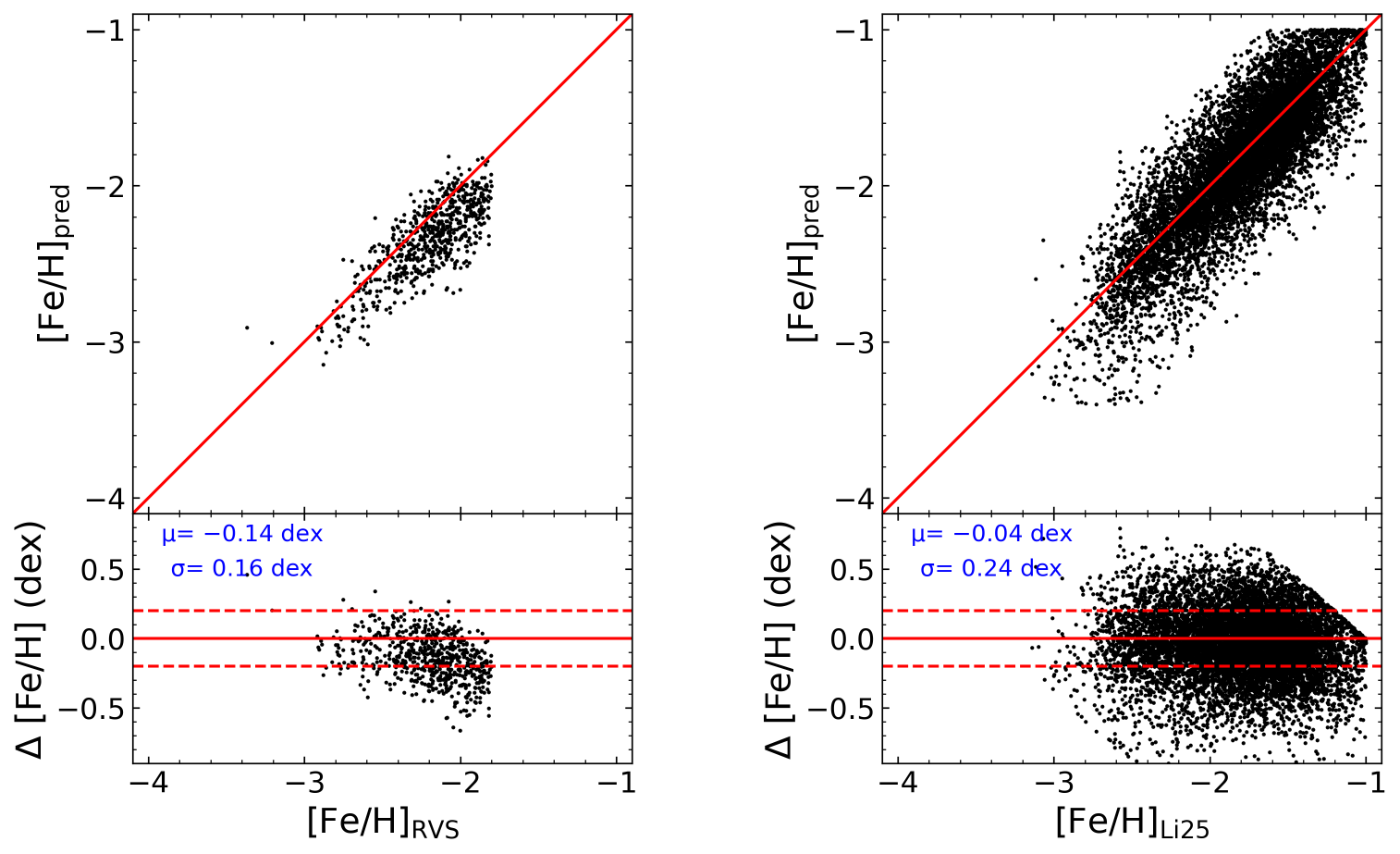}
\caption{Similar to Figure~\ref{fig6} but for the external validation of the metallicity for the giant stars using the sample from the Gaia RVS spectra (left panel) and \citet{2025ApJS..279...53L} (right panel).
\label{fig10}}
\end{figure*}

\begin{figure*}
\centering
\plotone{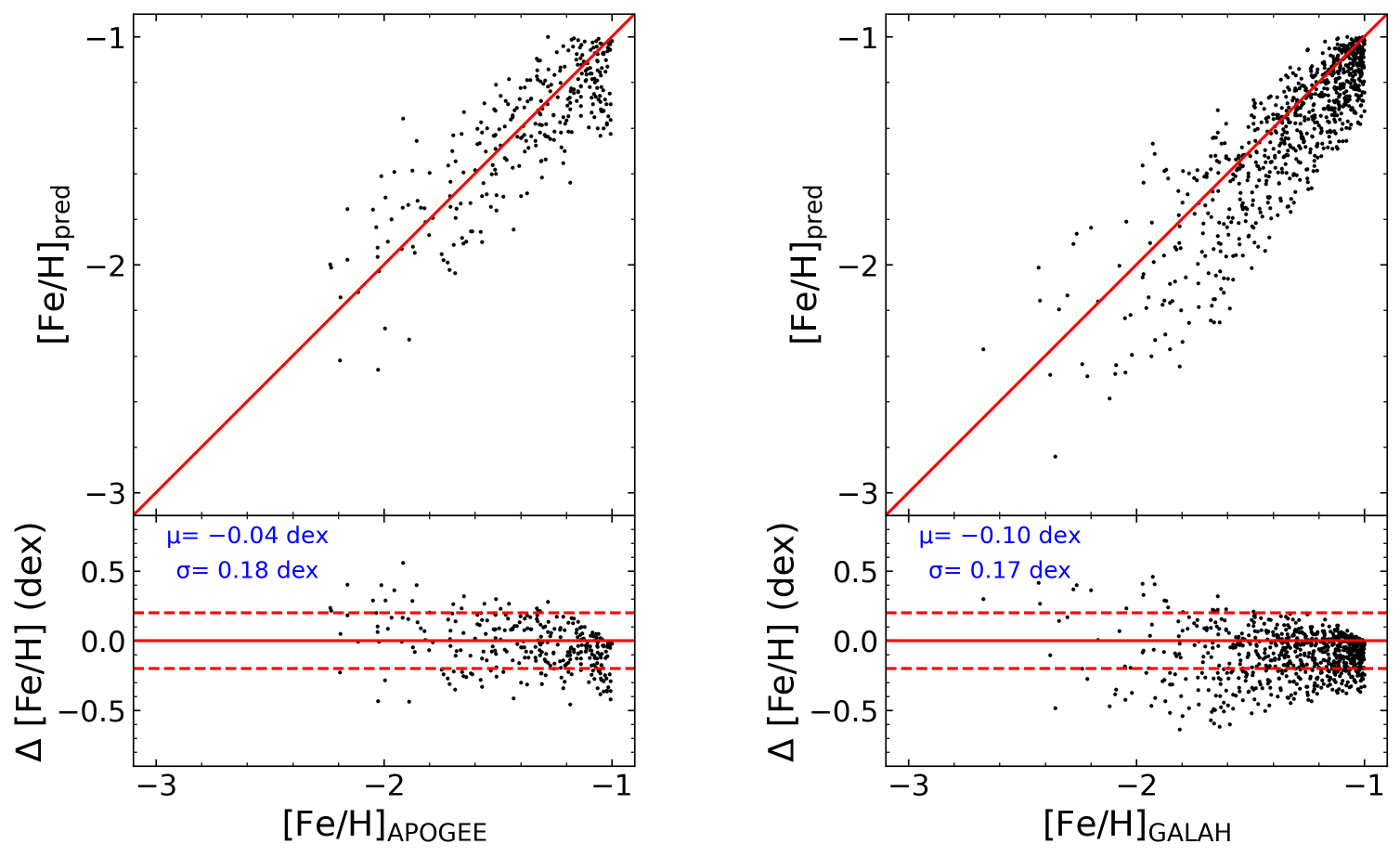}
\caption{Similar to Figure~\ref{fig6} but for the external validations of the metallicity for the dwarf stars using APOGEE DR17 (left panel) and GALAH DR4 (right panel).
\label{fig11}}
\end{figure*}

\begin{figure}
\centering
\plotone{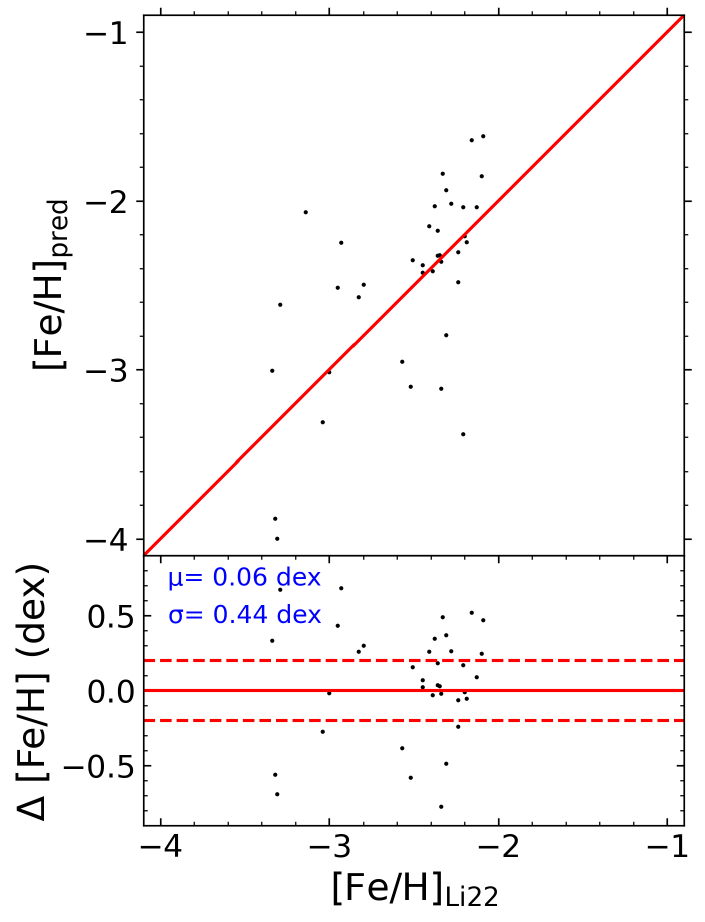}
\caption{Similar to Figure~\ref{fig6} but for the external validation of the metallicity for the dwarf stars using the catalog of \citet{2022ApJ...931..147L} .
\label{fig12}}
\end{figure}

\subsubsection{Validation for giant stars}

We first assess the performance of our method for giant stars using six independent spectroscopic samples, summarized in Table~\ref{Tab5}. 
The first comparison is carried out with the Apache Point Observatory Galactic Evolution Experiment (APOGEE; \citealt{2017AJ....154...94M}) DR17, which provides atmospheric parameters for more than 0.7 million stars. 
After cross-matching with Gaia XP spectra, we computed the synthetic $G_{\rm BP}$, $G_{\rm RP}$, and designed-filter $m$ magnitudes, and derive photometric metallicities using our fitted relations. 
Figure~\ref{fig7} presents the comparison between APOGEE [Fe/H] and our predicted values for giant stars, showing excellent agreement with a scatter of 0.15~dex. 
A similar comparison with GALAH DR4 (\citealt{2025PASA...42...51B}) yields a scatter of 0.17~dex. 
These results are consistent with the internal validation listed in Table~\ref{Tab4}, noting that most stars lie in the metallicity range $-2.5 < {\rm [Fe/H]} < -1.0$. 

Because APOGEE and GALAH mainly cover relatively metal-rich regimes ([Fe/H] $\gtrsim -2.5$), an additional very metal-poor sample is required to probe the low-metallicity end. 
We therefore adopt the high-resolution Subaru sample from \citet{2022ApJ...931..147L}, which contains $\sim$400 very metal-poor stars with accurate atmospheric parameters. 
Figure~8 shows the comparison between our predicted [Fe/H] and literature values, demonstrating good consistency with a scatter of 0.25~dex, again consistent with the internal validation. 
These results indicate that our method remains reliable down to [Fe/H] $\lesssim -3.0$.

As an additional independent test, we use member stars of globular clusters, which are expected to share nearly identical metallicities due to their common formation history. 
Cluster members are selected using the following criteria:
(1) positions within $10$ half-light radii ($r_{h}$; \citealt{2010arXiv1012.3224H}) from the cluster center; and 
(2) proper motions satisfying 
$|{\mu_{\alpha}-\mu_{\alpha,\rm GC}}| \le 8\sigma_{\mu_{\alpha,\rm GC}}$ and 
$|{\mu_{\delta}-\mu_{\delta,\rm GC}}| \le 8\sigma_{\mu_{\delta,\rm GC}}$, 
where cluster proper motions and uncertainties are adopted from \citet{2021MNRAS.505.5978V}. 
After cross-matching with Gaia XP spectra, metallicities are derived using our method. 
The final cluster sample is listed in Table~\ref{Tab6}, and Figure~\ref{fig9} shows the comparison with literature metallicities, yielding a scatter of 0.06~dex. 
The reported precision reflects the average metallicity of each globular cluster, rather than individual stars. If metallicities of all member stars were treated as independent measurements, the resulting scatter would increase substantially, to approximately 0.20~dex.

We further validate our results utilizing metallicities from Gaia RVS spectra \citep{2024A&A...683L..11V} and LAMOST low-resolution spectra \citep{2025ApJS..279...53L}. 
Both datasets estimate metallicity from the near-infrared Ca\,{\sc ii} triplet features, whose equivalent widths trace [Fe/H] down to $\sim-4.0$. 
As illustrated in Figure~\ref{fig10}, excellent agreement is achieved, with scatters of 0.16~dex and 0.24~dex, respectively. 
Overall, the precision of our photometric metallicities for giant stars is better than $\sim0.25$~dex across a wide metallicity range, demonstrating the robustness of the method.

\subsubsection{Validation for dwarf stars}

External validation for dwarf stars is more limited due to the scarcity of metal-poor dwarf samples. 
We use three spectroscopic datasets: APOGEE DR17, GALAH DR4, and the Subaru high-resolution sample from \citet{2022ApJ...931..147L}, as listed in Table~\ref{Tab7}. 
Nevertheless, the agreement remains good. 
The comparison with APOGEE shows a scatter of 0.18~dex (Figure~\ref{fig11}), while GALAH DR4 yields a scatter of 0.17~dex. 
For the extremely metal-poor dwarfs from \citet{2022ApJ...931..147L}, the scatter increases to 0.44~dex (Figure~\ref{fig12}), which is expected given the smaller sample size and the intrinsic difficulty of photometric metallicity estimation in this regime.

In summary, the external validations confirm that our method provides reliable metallicity estimates for both giant and dwarf stars, with typical precisions of $\sim0.15$--$0.25$~dex for giants and $\sim0.17$--$0.44$~dex for dwarfs over the metallicity range $-4 \lesssim {\rm [Fe/H]} \lesssim -1$.

\begin{table*}
\begin{center}
\caption{ The precision of the metallicity for giant  stars from external validations.}\label{Tab5}
\begin{tabular}{clccc}
  \hline\noalign{\smallskip}
   External samples  & $R$   & $N$     & Metallicity Range    & $\sigma_{\rm [Fe/H]}$     \\
     &   &   &    & (dex)   \\
  \hline\noalign{\smallskip}
   APOGEE DR17 & 22,500  & 7055  &    $-2.5 \le \rm[Fe/H] \le -1.0$    & 0.15      \\
   GALAH DR4 & 28,000  & 7163  &    $-3.4 \le \rm[Fe/H] \le -1.0$    & 0.17      \\
  \citet{2022ApJ...931..147L}  & 36,000  & 104  &    $-4.2 \le \rm[Fe/H] \le -2.0$    & 0.25      \\
   Golobular cluetsrs & --  &  10 &    $-2.4 \le \rm[Fe/H] \le -1.1$    & 0.20     \\
      Gaia RVS & 11,500  & 629  &    $-3.4 \le \rm[Fe/H] \le -1.5$    & 0.16      \\
  \citet{2025ApJS..279...53L}   & 1800  & 9551  &    $-3.1 \le \rm[Fe/H] \le -1.0$    & 0.24      \\

  \noalign{\smallskip}\hline
\end{tabular}
\end{center}
\tablecomments{Here $R$ represents the resolving power of the spectra, $N$ is the number of the stars/clusters from the compared external samples. Metallicity range is the validation range for our predicted metallicity, $\sigma_{\rm[Fe/H]}$ is the precision of the predicted metallicity for the giant stars.}
\end{table*}

\begin{table*}
\begin{center}
\caption{ Comparison of the photometric metallicties for globular clusters with values from \citet{2010arXiv1012.3224H}}\label{Tab6}
\begin{tabular}{cccccccccccc}
  \hline\noalign{\smallskip}
Name    &  $\rm [Fe/H]_{H10}$    &  ${\rm [Fe/H]_{TW}}$    & $\sigma_{\rm [Fe/H]_{TW}}$       &  $\rm [Fe/H]_{A23}$    
&  $\sigma_{\rm [Fe/H]_{A23}}$     &  $\rm [Fe/H]_{M24}$ 
&  $\sigma_{\rm [Fe/H]_{M24}}$    &  $\rm [Fe/H]_{Y25}$ 
&  $\sigma_{\rm [Fe/H]_{Y25}}$    &    N  \\
 &   (dex)   &  (dex)    & (dex) &   (dex) &  (dex)    & (dex)  &  (dex)    & (dex)   &  (dex)   &  \\
  \hline\noalign{\smallskip}
NGC6723             &   $-$1.10            &  $-$1.17    & 0.11 & $-$1.13     & 0.29   & $-$1.29   & 0.54   & $-$0.89   & 0.34    &  25      \\
NGC6981             &   $-$1.42            &  $-$1.25    & 0.03  & $-$1.18      & 0.09       & $-$1.69  & 0.15  & $-$1.22   & 0.09     &  6  \\
NGC5272             &   $-$1.50            &  $-$1.36    & 0.17  & $-$1.36       & 0.20   & $-$1.62  & 0.24   & $-$1.51  & 0.22    &   169      \\
NGC6205             &   $-$1.53           &  $-$1.45    & 0.19   & $-$1.30     &  0.31     &$-$1.61   & 0.22   &$-$1.61   & 0.23       &  141   \\
NGC6752             &   $-$1.54            &  $-$1.43    & 0.10  & $-$1.25     &  0.37      & $-$1.60  & 0.27    & $-$1.52  & 0.35    &  88    \\
NGC1904             &  $-$1.60             &  $-$1.55    & 0.24  & $-$1.43     &  0.23    & $-$1.69  &0.31   & $-$1.56  &0.29       & 83    \\
NGC6681             &  $-$1.62             &  $-$1.54    & 0.10  & $-$1.23     &  0.14    & $-$1.55  &0.09   & $-$1.37  &0.10     &  8     \\
NGC7089             &   $-$1.65            &  $-$1.52    & 0.23 & $-$1.34       & 0.28    & $-$1.59   &0.33   & $-$1.45   &0.40  & 89       \\
NGC6341             &    $-$2.31           &  $-$2.04    & 0.24  & $-$1.92      & 0.37   &$-$2.37  &0.21   &$-$2.48  &0.26        &   84  \\
NGC7078             &    $-$2.37           &  $-$2.22    & 0.17  & $-$2.15      & 0.34   &$-$2.36  &0.20    &$-$2.48  &0.28       &  125      \\
  \noalign{\smallskip}\hline
\end{tabular}
\end{center}
\tablecomments{Here $\rm[Fe/H]_{H10}$ denotes the metallicity from \citet{2010arXiv1012.3224H}. 
${\rm [Fe/H]_{TW}}$ and $\sigma_{\rm [Fe/H]_{TW}}$ represent the median metallicity and its scatter 
for the cluster member stars predicted by our method, respectively. 
$\rm[Fe/H]_{A23}$ and $\sigma_{\rm [Fe/H]_{A23}}$ give the median metallicity and scatter of the 
cluster member stars from \citet{2023ApJS..267....8A}. 
$\rm[Fe/H]_{M24}$ and $\sigma_{\rm [Fe/H]_{M24}}$ correspond to the median metallicity and scatter of the 
cluster member stars from \citet{2024A&A...692A.115M}. 
$\rm[Fe/H]_{Y25}$ and $\sigma_{\rm [Fe/H]_{Y25}}$ denote the median metallicity and scatter of the 
cluster member stars from \citet{2025ApJS..279....7Y}.}
\end{table*}

\begin{table*}
\begin{center}
\caption{ The precision of the metallicity for dwarf  stars from external validations.}\label{Tab7}
\begin{tabular}{clccc}
  \hline\noalign{\smallskip}
   External samples  & $R$   & $N$     & Metallicity Range    & $\sigma_{\rm [Fe/H]}$     \\
     &   &   & (dex)   & (dex)    \\
  \hline\noalign{\smallskip}
   APOGEE DR17 & 22,500  & 292  &    $-2.2 \le \rm[Fe/H] \le -1.0$    & 0.18      \\
   GALAH DR4 & 28,000  & 753  &    $-2.7 \le \rm[Fe/H] \le -1.0$    & 0.17      \\
  \citet{2022ApJ...931..147L}  & 36,000  & 38  &    $-3.3 \le \rm[Fe/H] \le -2.0$    & 0.44      \\

  \noalign{\smallskip}\hline
\end{tabular}
\end{center}
\tablecomments{Here $R$ represents the resolving power of the spectra, $N$ is the number of the stars from the compared external samples. Metallicity range is the validatoin range for our predicted metallicity, $\sigma_{\rm[Fe/H]}$ is the precision of the predicted metallicity for the dwarf stars.}
\end{table*}

\begin{table}
\begin{center}
\caption{ The number of the giant and dwarf stars in the final catalog.}\label{Tab8}
\begin{tabular}{clccc}
  \hline\noalign{\smallskip}
 $\sigma_{\rm [Fe/H]}$ (dex)  & Giant stars    & Dwarf stars \\
  \hline\noalign{\smallskip}
$ \le 0.3$    &  305,687    & 784,255   \\
$ \le 0.5$     & 1,247,796    & 2,489,975    \\
$ \le 1.0$     & 2,966,438    & 6,943,884    \\ 
  \noalign{\smallskip}\hline
\end{tabular}
\end{center}
\tablecomments{Here $\sigma_{\rm [Fe/H]}$ is the error of the predicted metallicity.}
\end{table}

\begin{table*}
\begin{center}
\caption{ Description of the final catalog.}\label{Tab9}
\begin{tabular}{clccccccc}
  \hline\noalign{\smallskip}
 Field   & Description   & Unit  \\
  \hline\noalign{\smallskip}
  ra   & R.A. from Gaia EDR3  & deg \\
  dec   & Decl. from Gaia EDR3  & deg \\
  rgeo   & Distance from \citet{2021AJ....161..147B}  & pc \\
  ebv   & Values of $E(B-V)$ from \citet{2025ApJS..280...15W}  & ...\\
  $\rm bp\_rp$   & Dereddened color ($G_{\rm BP}-G_{\rm RP} $) & ... \\
  $\rm m\_bp$   & Dereddened color ($m-G_{\rm BP}$)   & ... \\
  feh   & Photometric metallicity  & ... \\
  $\rm feh\_err$    & Uncertainty of photometric metallicity  & dex \\
    flag   & Flag to indicate the source is a dwarf or giant star  & ... \\
  \noalign{\smallskip}\hline
\end{tabular}
\end{center}
\end{table*}

\begin{figure*}
\centering
\plotone{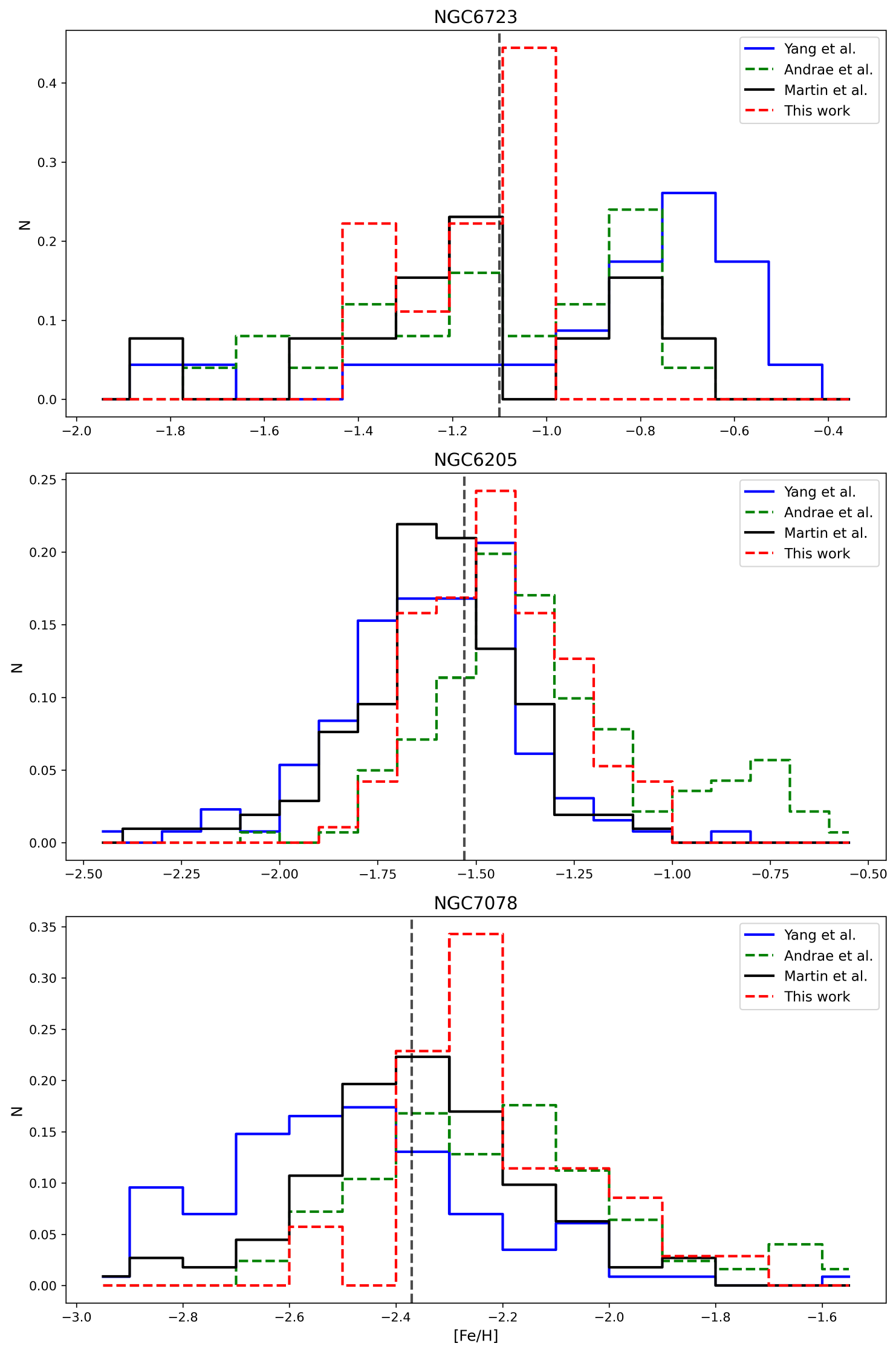}
\caption{The normalized number distribution of the member stars from NGC~6723, NGC~6205, NGC~7078, with red for this work, green for \citet{2023ApJS..267....8A}, blue for \citet{2025ApJS..279....7Y}, black for \citet{2024A&A...692A.115M}. The metallicity from H10 is marked by the dashed black line in each panel.
\label{fig13}}
\end{figure*}

\begin{figure*}
\centering
\plotone{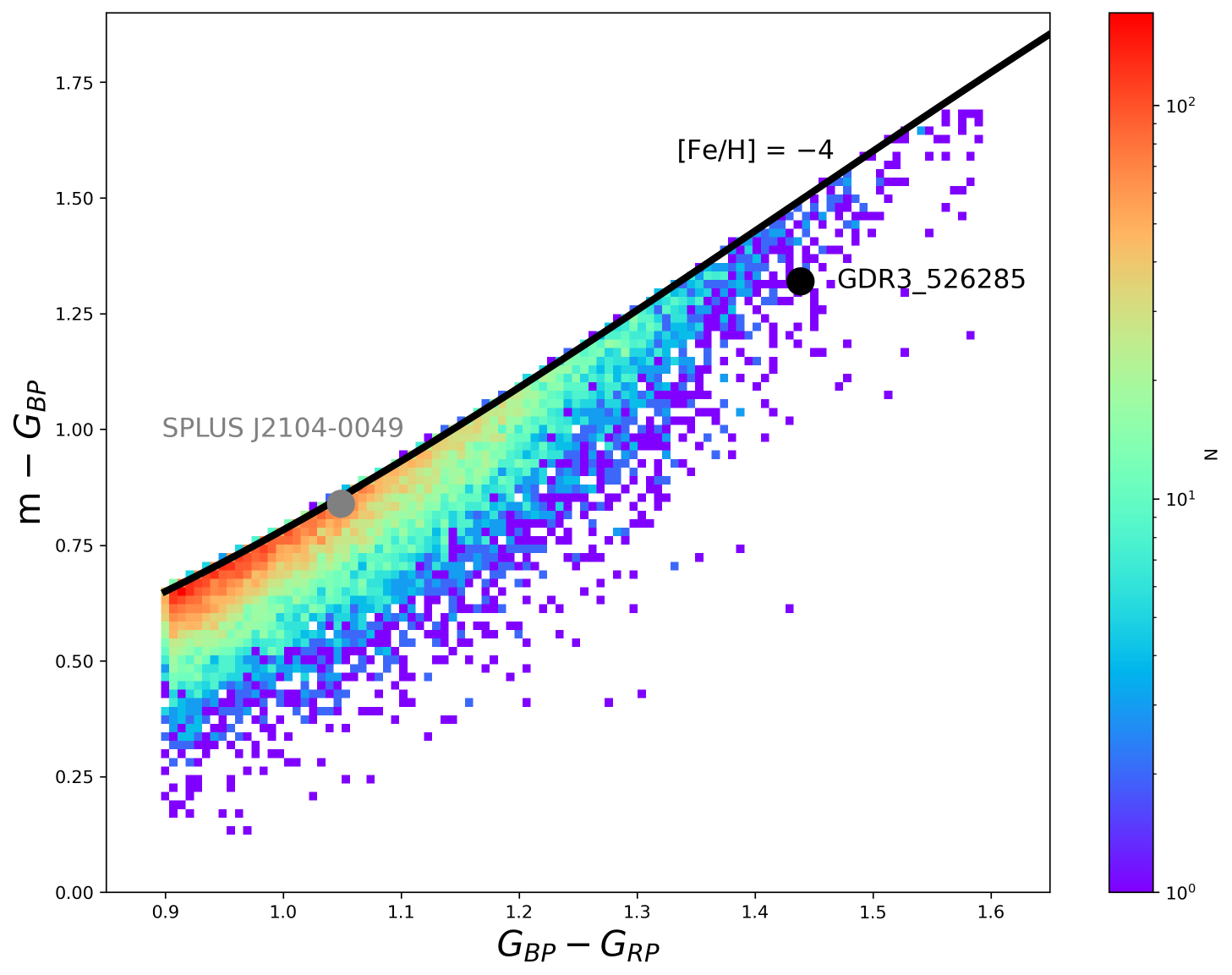}
\caption{Number density distribution of the red giant ultra metal-poor (UMP) candidates in the $(m - G_{\rm BP})$ vs. $(G_{\rm BP} - G_{\rm RP})$ color-color plane. The solid black line indicates the $\rm [Fe/H] = -4$ selection threshold for red giant stars, which is defined by Equation~3. Two spectroscopically confirmed red giant UMP stars from the literature are overplotted for comparison: the gray and black stars denote SPLUS J2104-0049 ($\rm [Fe/H] = -4.03$; \citealt{2021ApJ...912L..32P}) and GDR3\_526285 ($\rm [Fe/H] = -4.8$; \citealt{2025ApJ...989L..18L, 2025arXiv250921643J}), respectively.
\label{fig14}}
\end{figure*}

\section{Discussion} \label{sec:discussion}
The primary goal of this work is to optimize a filter in the near-ultraviolet (near-UV) band to achieve high-precision $\rm [Fe/H]$ measurements. Our methodology builds upon the framework described by \citet{2024ApJ...968L..24X}; however, while their study focuses mainly on stars with $\rm [Fe/H] \ge -1$, this work targets the metal-poor region ($\rm [Fe/H] \le -1$). These stars, formed during the early stages of the Milky Way, are essential for deciphering the Galaxy's early formation and chemical evolution. Our approach yields precise $\rm [Fe/H]$ estimates for metal-poor stars, extending down to $\rm [Fe/H] \approx -4$. Internal and external validations demonstrate that the predicted metallicities achieve high precision for giant stars, with a scatter smaller than 0.25~dex. For dwarf stars, while the precision is relatively lower, the scatter remains within 0.44~dex. Overall, these validations underscore the robustness of our method in characterizing the metallicity of metal-poor populations. However, the predicted metallicities show a systematic offset toward lower values compared to those from both APOGEE and GALAH, as shown in Figure~\ref{fig7}. To further investigate this trend, we compare metallicities for common sources between the original SAGA/PASTEL high-resolution spectroscopic (HRS) sample and GALAH DR4. The comparison, performed separately for dwarf and giant stars, is shown in Figure~\ref{fig_app_result}. Overall, the two datasets are broadly consistent. However, at the metal-poor end, the SAGA/PASTEL HRS metallicities tend to be systematically lower than those from GALAH DR4, particularly for giant stars. 
As mentioned above, the metallicities in the SAGA/PASTEL compilation are collected from multiple literature sources. One possible origin of this systematic offset is that most of these measurements are based on LTE analyses, whereas GALAH DR4 adopts NLTE corrections. 
We also note that part of the systematic offset at the metal-poor end may be influenced by boundary effects in the comparison datasets, as the metallicity coverage of GALAH DR4 and APOGEE is limited to $\mathrm{[Fe/H]} \sim -3$ and $\sim -2.5$, respectively.
To account for this systematic difference, we derive a two-order polynomial correction, as shown in the bottom panel of Figure~\ref{fig_app_result}, which enables transformation of our metallicity scale to that of GALAH DR4. This correction should be applied with caution and only within the calibrated metallicity range.

To further assess the performance of our method, we analyze member stars from three representative globular clusters (GCs) spanning a wide metallicity range: the relatively metal-rich NGC~6723 ($\rm [Fe/H] = -1.1$), the intermediate-metallicity NGC~6205 ($\rm [Fe/H] = -1.53$), and the metal-poor NGC~7078 ($\rm [Fe/H] = -2.37$). These clusters contain a significant number of giant member stars with available Gaia XP spectra, as detailed in Table~\ref{Tab5}. In Figure~\ref{fig13}, we present the $\rm [Fe/H]$ distributions for these member stars derived using our method and compare them with the results from \citet{2023ApJS..267....8A}, \citet{2024A&A...692A.115M} and \citet{2025ApJS..279....7Y}. Our results show that our measurements, both in terms of median values and dispersion, are in better agreement with literature benchmarks and exhibit the smallest scatter. This comparison highlights the superior capability of our method in measuring stellar metallicity, particularly at the metal-poor end.

Based on these validation results, we utilize the fitting coefficients listed in Table~\ref{Tab1} to process the entire Gaia XP sample. The stars are first classified into giants and dwarfs based on their positions in the color-absolute magnitude diagram (CMD). The dereddened colors $(m - G_{\rm BP})$ and $G_{\rm BP} - G_{\rm RP}$ are derived using the same method described in Section~2.1. The effective temperature adopted to determine the temperature- and reddening-dependent extinction coefficients is taken from \citet{2023MNRAS.524.1855Z} for the full sample. The metallicities of giant and dwarf stars, within their respective applicable color ranges, are then determined separately using a maximum likelihood approach. In this procedure, the metallicity grid for giant stars spans $-4.0 \leq \mathrm{[Fe/H]} \leq -1.0$ with a step of 0.01,dex, while for dwarf stars it covers $-3.3 \leq \mathrm{[Fe/H]} \leq -1.0$ with the same step size. Finallly, we have constructed a final catalog containing 14,536,547 metal-poor stars (comprising 4,074,728 giants and 10,461,819 dwarfs), providing both $\rm [Fe/H]$ values and their associated uncertainties (see Table~\ref{Tab8}). This catalog identifies approximately 1.35~million very metal-poor (VMP) star candidates and 0.16~million extremely metal-poor (EMP) star candidates. Table~\ref{Tab9} lists the columns included in the final catalog. Users can select subsamples with different levels of precision by applying cuts on the metallicity uncertainty, depending on their scientific goals. As illustrated in Section~2.3, the filter configuration adopted in this work is very similar to that of the Pristine Ca H\&K band. However, our design achieves higher metallicity sensitivity, leading to more precise metallicity estimates. Compared to previous studies, our work provides among the highest precision in metallicity determination (see Table~\ref{Tab6} and Figure~\ref{fig13}).

Furthermore, we provide a catalog of 19,923 red giant ultra-metal-poor (UMP) candidates\footnote{This table is available upon request.}. Figure~\ref{fig14} illustrates the distribution of these candidates. Notably, two red giant UMP giant stars in our sample have been confirmed by previous studies, i.e., SPLUS J2104$-$0049 with $\mathrm{[Fe/H]} = -4.03$ from \citet{2021ApJ...912L..32P}, and GDR3 526285 with $\mathrm{[Fe/H]} = -4.8$ from \citet{2025ApJ...989L..18L} and \citet{2025arXiv250921643J}. The consistency with independent high-resolution spectroscopic measurements highlights the potential of our catalog for efficiently identifying and discovering new UMP stars. However, the metallicities of these two stars cannot be recovered by our method, as they fall below the lower limit of $\mathrm{[Fe/H]} = -4$ imposed by Equation~3. The full catalog is available on Zenodo at \href{https://doi.org/10.5281/zenodo.19176297}{10.5281/zenodo.19176298}.

\section{Summary} \label{sec:summary}
In summary, we have determined the optimal filter configurations for estimating the stellar metallicity of metal-poor stars. The optimized parameters are found to be $\lambda_{\rm c} = 3960$\,\AA\ and $\Delta\lambda = 80$\,\AA\ for giant stars, and $\lambda_{\rm c} = 3920$\,\AA\ and $\Delta\lambda = 80$\,\AA\ for dwarf stars. Following classification via the CMD, metallicities for both populations are derived independently using a stellar loci fitting method based on these optimized filters. Both internal and external validations demonstrate that our predicted metallicities are in excellent agreement with the literature values. Consequently, we present a large-scale catalog of approximately 14.5 million metal-poor stars derived from Gaia XP spectra. This sample comprises 4.0 million giant stars within the color range $0.9 \le (G_{\rm BP}-G_{\rm RP}) \le 1.8$ and 10.5 million dwarf stars within $0.65 \le (G_{\rm BP}-G_{\rm RP}) \le 1.2$. According to the validation metallicity ranges, we caution that the reliable lower limit of the metallicity for giant and dwarf stars is $\rm [Fe/H] \approx -4$, $\rm [Fe/H] \approx -3.3$, respectively. This extensive catalog is expected to provide critical insights into the early formation and chemical evolution of the Milky Way.

\section*{acknowledgments}
Y.H. acknowledges the support from the National Science Foundation of China (NSFC grant No. 12422303), the Fundamental Research Funds for the Central Universities (grant Nos. 118900M122, E5EQ3301X2, and E4EQ3301X2), and the National Key R\&D Programme of China (grant No. 2023YFA1608303). 
This work is also supported by the National Key Basic R\&D Program of China via 2023YFA1608303 and the Strategic Priority Research Program of the Chinese Academy of Sciences (XDB0550103). C.J.Z. acknowledges support from China Postdoctoral Science Foundation (grant No. 2025M783229).

%% For this sample we use BibTeX plus aasjournals.bst to generate the
%% the bibliography. The sample631.bib file was populated from ADS. To
%% get the citations to show in the compiled file do the following:
%%
%% pdflatex sample631.tex
%% bibtext sample631
%% pdflatex sample631.tex
%% pdflatex sample631.tex

\bibliography{sample631}{}

@ARTICLE{2016ARA&A..54..529B,
       author = {{Bland-Hawthorn}, Joss and {Gerhard}, Ortwin},
        title = "{The Galaxy in Context: Structural, Kinematic, and Integrated Properties}",
      journal = {\araa},
         year = 2016,
        month = sep,
       volume = {54},
        pages = {529-596},
          doi = {10.1146/annurev-astro-081915-023441},
archivePrefix = {arXiv},
       eprint = {1602.07702},
 primaryClass = {astro-ph.GA},
       adsurl = {https://ui.adsabs.harvard.edu/abs/2016ARA&A..54..529B}
}

@ARTICLE{2018MNRAS.481.4093S,
       author = {{Sanders}, Jason L. and {Das}, Payel},
        title = "{Isochrone ages for {\ensuremath{\sim}}3 million stars with the second Gaia data release}",
      journal = {\mnras},
         year = 2018,
        month = dec,
       volume = {481},
       number = {3},
        pages = {4093-4110},
          doi = {10.1093/mnras/sty2490},
archivePrefix = {arXiv},
       eprint = {1806.02324},
 primaryClass = {astro-ph.GA},
       adsurl = {https://ui.adsabs.harvard.edu/abs/2018MNRAS.481.4093S}
}

@ARTICLE{2024RAA....24d5015S,
       author = {{Shi}, Rui-Feng and {Huang}, Yang and {Li}, Xin-Yi and {Zhang}, Hua-Wei},
        title = "{On the Performances of Estimating Stellar Atmospheric Parameters from CSST Broad-band Photometry}",
      journal = {Research in Astronomy and Astrophysics},
         year = 2024,
        month = apr,
       volume = {24},
       number = {4},
          eid = {045015},
        pages = {045015},
          doi = {10.1088/1674-4527/ad2dbd},
archivePrefix = {arXiv},
       eprint = {2401.05948},
 primaryClass = {astro-ph.SR},
       adsurl = {https://ui.adsabs.harvard.edu/abs/2024RAA....24d5015S}
}

@ARTICLE{2023A&A...674A..33G,
       author = {{Gaia Collaboration} and {Montegriffo}, P. and {Bellazzini}, M. and {De Angeli}, F. and {Andrae}, R. and {Barstow}, M.~A. and {Bossini}, D. and {Bragaglia}, A. and {Burgess}, P.~W. and {Cacciari}, C. and {Carrasco}, J.~M. and {Chornay}, N. and {Delchambre}, L. and {Evans}, D.~W. and {Fouesneau}, M. and {Fr{\'e}mat}, Y. and {Garabato}, D. and {Jordi}, C. and {Manteiga}, M. and {Massari}, D. and {Palaversa}, L. and {Pancino}, E. and {Riello}, M. and {Ruz Mieres}, D. and {Sanna}, N. and {Santove{\~n}a}, R. and {Sordo}, R. and {Vallenari}, A. and {Walton}, N.~A. and {Brown}, A.~G.~A. and {Prusti}, T. and {de Bruijne}, J.~H.~J. and {Arenou}, F. and {Babusiaux}, C. and {Biermann}, M. and {Creevey}, O.~L. and {Ducourant}, C. and {Eyer}, L. and {Guerra}, R. and {Hutton}, A. and {Klioner}, S.~A. and {Lammers}, U.~L. and {Lindegren}, L. and {Luri}, X. and {Mignard}, F. and {Panem}, C. and {Pourbaix}, D. and {Randich}, S. and {Sartoretti}, P. and {Soubiran}, C. and {Tanga}, P. and {Bailer-Jones}, C.~A.~L. and {Bastian}, U. and {Drimmel}, R. and {Jansen}, F. and {Katz}, D. and {Lattanzi}, M.~G. and {van Leeuwen}, F. and {Bakker}, J. and {Casta{\~n}eda}, J. and {Fabricius}, C. and {Galluccio}, L. and {Guerrier}, A. and {Heiter}, U. and {Masana}, E. and {Messineo}, R. and {Mowlavi}, N. and {Nicolas}, C. and {Nienartowicz}, K. and {Pailler}, F. and {Panuzzo}, P. and {Riclet}, F. and {Roux}, W. and {Seabroke}, G.~M. and {Th{\'e}venin}, F. and {Gracia-Abril}, G. and {Portell}, J. and {Teyssier}, D. and {Altmann}, M. and {Audard}, M. and {Bellas-Velidis}, I. and {Benson}, K. and {Berthier}, J. and {Blomme}, R. and {Busonero}, D. and {Busso}, G. and {C{\'a}novas}, H. and {Carry}, B. and {Cellino}, A. and {Cheek}, N. and {Clementini}, G. and {Damerdji}, Y. and {Davidson}, M. and {de Teodoro}, P. and {Nu{\~n}ez Campos}, M. and {Dell'Oro}, A. and {Esquej}, P. and {Fern{\'a}ndez-Hern{\'a}ndez}, J. and {Fraile}, E. and {Garc{\'\i}a-Lario}, P. and {Gosset}, E. and {Haigron}, R. and {Halbwachs}, J. -L. and {Hambly}, N.~C. and {Harrison}, D.~L. and {Hern{\'a}ndez}, J. and {Hestroffer}, D. and {Hodgkin}, S.~T. and {Holl}, B. and {Jan{\ss}en}, K. and {Jevardat de Fombelle}, G. and {Jordan}, S. and {Krone-Martins}, A. and {Lanzafame}, A.~C. and {L{\"o}ffler}, W. and {Marchal}, O. and {Marrese}, P.~M. and {Moitinho}, A. and {Muinonen}, K. and {Osborne}, P. and {Pauwels}, T. and {Recio-Blanco}, A. and {Reyl{\'e}}, C. and {Rimoldini}, L. and {Roegiers}, T. and {Rybizki}, J. and {Sarro}, L.~M. and {Siopis}, C. and {Smith}, M. and {Sozzetti}, A. and {Utrilla}, E. and {van Leeuwen}, M. and {Abbas}, U. and {{\'A}brah{\'a}m}, P. and {Abreu Aramburu}, A. and {Aerts}, C. and {Aguado}, J.~J. and {Ajaj}, M. and {Aldea-Montero}, F. and {Altavilla}, G. and {{\'A}lvarez}, M.~A. and {Alves}, J. and {Anderson}, R.~I. and {Anglada Varela}, E. and {Antoja}, T. and {Baines}, D. and {Baker}, S.~G. and {Balaguer-N{\'u}{\~n}ez}, L. and {Balbinot}, E. and {Balog}, Z. and {Barache}, C. and {Barbato}, D. and {Barros}, M. and {Bartolom{\'e}}, S. and {Bassilana}, J. -L. and {Bauchet}, N. and {Becciani}, U. and {Berihuete}, A. and {Bernet}, M. and {Bertone}, S. and {Bianchi}, L. and {Binnenfeld}, A. and {Blanco-Cuaresma}, S. and {Boch}, T. and {Bombrun}, A. and {Bouquillon}, S. and {Bramante}, L. and {Breedt}, E. and {Bressan}, A. and {Brouillet}, N. and {Brugaletta}, E. and {Bucciarelli}, B. and {Burlacu}, A. and {Butkevich}, A.~G. and {Buzzi}, R. and {Caffau}, E. and {Cancelliere}, R. and {Cantat-Gaudin}, T. and {Carballo}, R. and {Carlucci}, T. and {Carnerero}, M.~I. and {Casamiquela}, L. and {Castellani}, M. and {Castro-Ginard}, A. and {Chaoul}, L. and {Charlot}, P. and {Chemin}, L. and {Chiaramida}, V. and {Chiavassa}, A. and {Comoretto}, G. and {Contursi}, G. and {Cooper}, W.~J. and {Cornez}, T. and {Cowell}, S. and {Crifo}, F. and {Cropper}, M. and {Crosta}, M. and {Crowley}, C. and {Dafonte}, C. and {Dapergolas}, A.},
        title = "{Gaia Data Release 3. The Galaxy in your preferred colours: Synthetic photometry from Gaia low-resolution spectra}",
      journal = {\aap},
         year = 2023,
        month = jun,
       volume = {674},
          eid = {A33},
        pages = {A33},
          doi = {10.1051/0004-6361/202243709},
archivePrefix = {arXiv},
       eprint = {2206.06215},
 primaryClass = {astro-ph.SR},
       adsurl = {https://ui.adsabs.harvard.edu/abs/2023A&A...674A..33G}
}

@ARTICLE{2008PASJ...60.1159S,
       author = {{Suda}, Takuma and {Katsuta}, Yutaka and {Yamada}, Shimako and {Suwa}, Tamon and {Ishizuka}, Chikako and {Komiya}, Yutaka and {Sorai}, Kazuo and {Aikawa}, Masayuki and {Fujimoto}, Masayuki Y.},
        title = "{Stellar Abundances for the Galactic Archeology (SAGA) Database --- Compilation of the Characteristics of Known Extremely Metal-Poor Stars}",
      journal = {\pasj},
         year = 2008,
        month = oct,
       volume = {60},
        pages = {1159},
          doi = {10.1093/pasj/60.5.1159},
archivePrefix = {arXiv},
       eprint = {0806.3697},
 primaryClass = {astro-ph},
       adsurl = {https://ui.adsabs.harvard.edu/abs/2008PASJ...60.1159S}
}

@ARTICLE{2022A&A...663A...4S,
       author = {{Soubiran}, C. and {Brouillet}, N. and {Casamiquela}, L.},
        title = "{Assessment of [Fe/H] determinations for FGK stars in spectroscopic surveys}",
      journal = {\aap},
         year = 2022,
        month = jul,
       volume = {663},
          eid = {A4},
        pages = {A4},
          doi = {10.1051/0004-6361/202142409},
archivePrefix = {arXiv},
       eprint = {2112.07545},
 primaryClass = {astro-ph.SR},
       adsurl = {https://ui.adsabs.harvard.edu/abs/2022A&A...663A...4S}
}

@ARTICLE{2022ApJ...931..147L,
       author = {{Li}, Haining and {Aoki}, Wako and {Matsuno}, Tadafumi and {Xing}, Qianfan and {Suda}, Takuma and {Tominaga}, Nozomu and {Chen}, Yuqin and {Honda}, Satoshi and {Ishigaki}, Miho N. and {Shi}, Jianrong and {Zhao}, Jingkun and {Zhao}, Gang},
        title = "{Four-hundred Very Metal-poor Stars Studied with LAMOST and Subaru. II. Elemental Abundances}",
      journal = {\apj},
         year = 2022,
        month = jun,
       volume = {931},
       number = {2},
          eid = {147},
        pages = {147},
          doi = {10.3847/1538-4357/ac6514},
archivePrefix = {arXiv},
       eprint = {2203.11529},
 primaryClass = {astro-ph.SR},
       adsurl = {https://ui.adsabs.harvard.edu/abs/2022ApJ...931..147L}
}

@ARTICLE{2010arXiv1012.3224H,
       author = {{Harris}, William E.},
        title = "{A New Catalog of Globular Clusters in the Milky Way}",
      journal = {arXiv e-prints},
         year = 2010,
        month = dec,
          eid = {arXiv:1012.3224},
        pages = {arXiv:1012.3224},
          doi = {10.48550/arXiv.1012.3224},
archivePrefix = {arXiv},
       eprint = {1012.3224},
 primaryClass = {astro-ph.GA},
       adsurl = {https://ui.adsabs.harvard.edu/abs/2010arXiv1012.3224H}
}

@ARTICLE{2021MNRAS.505.5978V,
       author = {{Vasiliev}, Eugene and {Baumgardt}, Holger},
        title = "{Gaia EDR3 view on galactic globular clusters}",
      journal = {\mnras},
         year = 2021,
        month = aug,
       volume = {505},
       number = {4},
        pages = {5978-6002},
          doi = {10.1093/mnras/stab1475},
archivePrefix = {arXiv},
       eprint = {2102.09568},
 primaryClass = {astro-ph.GA},
       adsurl = {https://ui.adsabs.harvard.edu/abs/2021MNRAS.505.5978V}
}

@ARTICLE{2024ApJ...968L..24X,
       author = {{Xiao}, Kai and {Huang}, Bowen and {Huang}, Yang and {Yuan}, Haibo and {Beers}, Timothy C. and {Liu}, Jifeng and {Xiang}, Maosheng and {Lu}, Xue and {Xu}, Shuai and {Yang}, Lin and {Zheng}, Chuanjie and {Li}, Zhirui and {Zhang}, Bowen and {Shi}, Ruifeng},
        title = "{Filter Design for Estimation of Stellar Metallicity: Insights from Experiments with Gaia XP Spectra}",
      journal = {\apjl},
         year = 2024,
        month = jun,
       volume = {968},
       number = {2},
          eid = {L24},
        pages = {L24},
          doi = {10.3847/2041-8213/ad5205},
archivePrefix = {arXiv},
       eprint = {2405.20212},
 primaryClass = {astro-ph.SR},
       adsurl = {https://ui.adsabs.harvard.edu/abs/2024ApJ...968L..24X}
}

@ARTICLE{2017AJ....154...94M,
       author = {{Majewski}, Steven R. and {Schiavon}, Ricardo P. and {Frinchaboy}, Peter M. and {Allende Prieto}, Carlos and {Barkhouser}, Robert and {Bizyaev}, Dmitry and {Blank}, Basil and {Brunner}, Sophia and {Burton}, Adam and {Carrera}, Ricardo and {Chojnowski}, S. Drew and {Cunha}, K{\'a}tia and {Epstein}, Courtney and {Fitzgerald}, Greg and {Garc{\'\i}a P{\'e}rez}, Ana E. and {Hearty}, Fred R. and {Henderson}, Chuck and {Holtzman}, Jon A. and {Johnson}, Jennifer A. and {Lam}, Charles R. and {Lawler}, James E. and {Maseman}, Paul and {M{\'e}sz{\'a}ros}, Szabolcs and {Nelson}, Matthew and {Nguyen}, Duy Coung and {Nidever}, David L. and {Pinsonneault}, Marc and {Shetrone}, Matthew and {Smee}, Stephen and {Smith}, Verne V. and {Stolberg}, Todd and {Skrutskie}, Michael F. and {Walker}, Eric and {Wilson}, John C. and {Zasowski}, Gail and {Anders}, Friedrich and {Basu}, Sarbani and {Beland}, Stephane and {Blanton}, Michael R. and {Bovy}, Jo and {Brownstein}, Joel R. and {Carlberg}, Joleen and {Chaplin}, William and {Chiappini}, Cristina and {Eisenstein}, Daniel J. and {Elsworth}, Yvonne and {Feuillet}, Diane and {Fleming}, Scott W. and {Galbraith-Frew}, Jessica and {Garc{\'\i}a}, Rafael A. and {Garc{\'\i}a-Hern{\'a}ndez}, D. An{\'\i}bal and {Gillespie}, Bruce A. and {Girardi}, L{\'e}o and {Gunn}, James E. and {Hasselquist}, Sten and {Hayden}, Michael R. and {Hekker}, Saskia and {Ivans}, Inese and {Kinemuchi}, Karen and {Klaene}, Mark and {Mahadevan}, Suvrath and {Mathur}, Savita and {Mosser}, Beno{\^\i}t and {Muna}, Demitri and {Munn}, Jeffrey A. and {Nichol}, Robert C. and {O'Connell}, Robert W. and {Parejko}, John K. and {Robin}, A.~C. and {Rocha-Pinto}, Helio and {Schultheis}, Matthias and {Serenelli}, Aldo M. and {Shane}, Neville and {Silva Aguirre}, Victor and {Sobeck}, Jennifer S. and {Thompson}, Benjamin and {Troup}, Nicholas W. and {Weinberg}, David H. and {Zamora}, Olga},
        title = "{The Apache Point Observatory Galactic Evolution Experiment (APOGEE)}",
      journal = {\aj},
         year = 2017,
        month = sep,
       volume = {154},
       number = {3},
          eid = {94},
        pages = {94},
          doi = {10.3847/1538-3881/aa784d},
archivePrefix = {arXiv},
       eprint = {1509.05420},
 primaryClass = {astro-ph.IM},
       adsurl = {https://ui.adsabs.harvard.edu/abs/2017AJ....154...94M}
}

@ARTICLE{2016A&A...595A...1G,
       author = {{Gaia Collaboration} and {Prusti}, T. and {de Bruijne}, J.~H.~J. and {Brown}, A.~G.~A. and {Vallenari}, A. and {Babusiaux}, C. and {Bailer-Jones}, C.~A.~L. and {Bastian}, U. and {Biermann}, M. and {Evans}, D.~W. and {Eyer}, L. and {Jansen}, F. and {Jordi}, C. and {Klioner}, S.~A. and {Lammers}, U. and {Lindegren}, L. and {Luri}, X. and {Mignard}, F. and {Milligan}, D.~J. and {Panem}, C. and {Poinsignon}, V. and {Pourbaix}, D. and {Randich}, S. and {Sarri}, G. and {Sartoretti}, P. and {Siddiqui}, H.~I. and {Soubiran}, C. and {Valette}, V. and {van Leeuwen}, F. and {Walton}, N.~A. and {Aerts}, C. and {Arenou}, F. and {Cropper}, M. and {Drimmel}, R. and {H{\o}g}, E. and {Katz}, D. and {Lattanzi}, M.~G. and {O'Mullane}, W. and {Grebel}, E.~K. and {Holland}, A.~D. and {Huc}, C. and {Passot}, X. and {Bramante}, L. and {Cacciari}, C. and {Casta{\~n}eda}, J. and {Chaoul}, L. and {Cheek}, N. and {De Angeli}, F. and {Fabricius}, C. and {Guerra}, R. and {Hern{\'a}ndez}, J. and {Jean-Antoine-Piccolo}, A. and {Masana}, E. and {Messineo}, R. and {Mowlavi}, N. and {Nienartowicz}, K. and {Ord{\'o}{\~n}ez-Blanco}, D. and {Panuzzo}, P. and {Portell}, J. and {Richards}, P.~J. and {Riello}, M. and {Seabroke}, G.~M. and {Tanga}, P. and {Th{\'e}venin}, F. and {Torra}, J. and {Els}, S.~G. and {Gracia-Abril}, G. and {Comoretto}, G. and {Garcia-Reinaldos}, M. and {Lock}, T. and {Mercier}, E. and {Altmann}, M. and {Andrae}, R. and {Astraatmadja}, T.~L. and {Bellas-Velidis}, I. and {Benson}, K. and {Berthier}, J. and {Blomme}, R. and {Busso}, G. and {Carry}, B. and {Cellino}, A. and {Clementini}, G. and {Cowell}, S. and {Creevey}, O. and {Cuypers}, J. and {Davidson}, M. and {De Ridder}, J. and {de Torres}, A. and {Delchambre}, L. and {Dell'Oro}, A. and {Ducourant}, C. and {Fr{\'e}mat}, Y. and {Garc{\'\i}a-Torres}, M. and {Gosset}, E. and {Halbwachs}, J. -L. and {Hambly}, N.~C. and {Harrison}, D.~L. and {Hauser}, M. and {Hestroffer}, D. and {Hodgkin}, S.~T. and {Huckle}, H.~E. and {Hutton}, A. and {Jasniewicz}, G. and {Jordan}, S. and {Kontizas}, M. and {Korn}, A.~J. and {Lanzafame}, A.~C. and {Manteiga}, M. and {Moitinho}, A. and {Muinonen}, K. and {Osinde}, J. and {Pancino}, E. and {Pauwels}, T. and {Petit}, J. -M. and {Recio-Blanco}, A. and {Robin}, A.~C. and {Sarro}, L.~M. and {Siopis}, C. and {Smith}, M. and {Smith}, K.~W. and {Sozzetti}, A. and {Thuillot}, W. and {van Reeven}, W. and {Viala}, Y. and {Abbas}, U. and {Abreu Aramburu}, A. and {Accart}, S. and {Aguado}, J.~J. and {Allan}, P.~M. and {Allasia}, W. and {Altavilla}, G. and {{\'A}lvarez}, M.~A. and {Alves}, J. and {Anderson}, R.~I. and {Andrei}, A.~H. and {Anglada Varela}, E. and {Antiche}, E. and {Antoja}, T. and {Ant{\'o}n}, S. and {Arcay}, B. and {Atzei}, A. and {Ayache}, L. and {Bach}, N. and {Baker}, S.~G. and {Balaguer-N{\'u}{\~n}ez}, L. and {Barache}, C. and {Barata}, C. and {Barbier}, A. and {Barblan}, F. and {Baroni}, M. and {Barrado y Navascu{\'e}s}, D. and {Barros}, M. and {Barstow}, M.~A. and {Becciani}, U. and {Bellazzini}, M. and {Bellei}, G. and {Bello Garc{\'\i}a}, A. and {Belokurov}, V. and {Bendjoya}, P. and {Berihuete}, A. and {Bianchi}, L. and {Bienaym{\'e}}, O. and {Billebaud}, F. and {Blagorodnova}, N. and {Blanco-Cuaresma}, S. and {Boch}, T. and {Bombrun}, A. and {Borrachero}, R. and {Bouquillon}, S. and {Bourda}, G. and {Bouy}, H. and {Bragaglia}, A. and {Breddels}, M.~A. and {Brouillet}, N. and {Br{\"u}semeister}, T. and {Bucciarelli}, B. and {Budnik}, F. and {Burgess}, P. and {Burgon}, R. and {Burlacu}, A. and {Busonero}, D. and {Buzzi}, R. and {Caffau}, E. and {Cambras}, J. and {Campbell}, H. and {Cancelliere}, R. and {Cantat-Gaudin}, T. and {Carlucci}, T. and {Carrasco}, J.~M. and {Castellani}, M. and {Charlot}, P. and {Charnas}, J. and {Charvet}, P. and {Chassat}, F. and {Chiavassa}, A. and {Clotet}, M. and {Cocozza}, G. and {Collins}, R.~S. and {Collins}, P. and {Costigan}, G.},
        title = "{The Gaia mission}",
      journal = {\aap},
         year = 2016,
        month = nov,
       volume = {595},
          eid = {A1},
        pages = {A1},
          doi = {10.1051/0004-6361/201629272},
archivePrefix = {arXiv},
       eprint = {1609.04153},
 primaryClass = {astro-ph.IM},
       adsurl = {https://ui.adsabs.harvard.edu/abs/2016A&A...595A...1G}
}

@ARTICLE{2016A&A...595A...2G,
       author = {{Gaia Collaboration} and {Brown}, A.~G.~A. and {Vallenari}, A. and {Prusti}, T. and {de Bruijne}, J.~H.~J. and {Mignard}, F. and {Drimmel}, R. and {Babusiaux}, C. and {Bailer-Jones}, C.~A.~L. and {Bastian}, U. and {Biermann}, M. and {Evans}, D.~W. and {Eyer}, L. and {Jansen}, F. and {Jordi}, C. and {Katz}, D. and {Klioner}, S.~A. and {Lammers}, U. and {Lindegren}, L. and {Luri}, X. and {O'Mullane}, W. and {Panem}, C. and {Pourbaix}, D. and {Randich}, S. and {Sartoretti}, P. and {Siddiqui}, H.~I. and {Soubiran}, C. and {Valette}, V. and {van Leeuwen}, F. and {Walton}, N.~A. and {Aerts}, C. and {Arenou}, F. and {Cropper}, M. and {H{\o}g}, E. and {Lattanzi}, M.~G. and {Grebel}, E.~K. and {Holland}, A.~D. and {Huc}, C. and {Passot}, X. and {Perryman}, M. and {Bramante}, L. and {Cacciari}, C. and {Casta{\~n}eda}, J. and {Chaoul}, L. and {Cheek}, N. and {De Angeli}, F. and {Fabricius}, C. and {Guerra}, R. and {Hern{\'a}ndez}, J. and {Jean-Antoine-Piccolo}, A. and {Masana}, E. and {Messineo}, R. and {Mowlavi}, N. and {Nienartowicz}, K. and {Ord{\'o}{\~n}ez-Blanco}, D. and {Panuzzo}, P. and {Portell}, J. and {Richards}, P.~J. and {Riello}, M. and {Seabroke}, G.~M. and {Tanga}, P. and {Th{\'e}venin}, F. and {Torra}, J. and {Els}, S.~G. and {Gracia-Abril}, G. and {Comoretto}, G. and {Garcia-Reinaldos}, M. and {Lock}, T. and {Mercier}, E. and {Altmann}, M. and {Andrae}, R. and {Astraatmadja}, T.~L. and {Bellas-Velidis}, I. and {Benson}, K. and {Berthier}, J. and {Blomme}, R. and {Busso}, G. and {Carry}, B. and {Cellino}, A. and {Clementini}, G. and {Cowell}, S. and {Creevey}, O. and {Cuypers}, J. and {Davidson}, M. and {De Ridder}, J. and {de Torres}, A. and {Delchambre}, L. and {Dell'Oro}, A. and {Ducourant}, C. and {Fr{\'e}mat}, Y. and {Garc{\'\i}a-Torres}, M. and {Gosset}, E. and {Halbwachs}, J. -L. and {Hambly}, N.~C. and {Harrison}, D.~L. and {Hauser}, M. and {Hestroffer}, D. and {Hodgkin}, S.~T. and {Huckle}, H.~E. and {Hutton}, A. and {Jasniewicz}, G. and {Jordan}, S. and {Kontizas}, M. and {Korn}, A.~J. and {Lanzafame}, A.~C. and {Manteiga}, M. and {Moitinho}, A. and {Muinonen}, K. and {Osinde}, J. and {Pancino}, E. and {Pauwels}, T. and {Petit}, J. -M. and {Recio-Blanco}, A. and {Robin}, A.~C. and {Sarro}, L.~M. and {Siopis}, C. and {Smith}, M. and {Smith}, K.~W. and {Sozzetti}, A. and {Thuillot}, W. and {van Reeven}, W. and {Viala}, Y. and {Abbas}, U. and {Abreu Aramburu}, A. and {Accart}, S. and {Aguado}, J.~J. and {Allan}, P.~M. and {Allasia}, W. and {Altavilla}, G. and {{\'A}lvarez}, M.~A. and {Alves}, J. and {Anderson}, R.~I. and {Andrei}, A.~H. and {Anglada Varela}, E. and {Antiche}, E. and {Antoja}, T. and {Ant{\'o}n}, S. and {Arcay}, B. and {Bach}, N. and {Baker}, S.~G. and {Balaguer-N{\'u}{\~n}ez}, L. and {Barache}, C. and {Barata}, C. and {Barbier}, A. and {Barblan}, F. and {Barrado y Navascu{\'e}s}, D. and {Barros}, M. and {Barstow}, M.~A. and {Becciani}, U. and {Bellazzini}, M. and {Bello Garc{\'\i}a}, A. and {Belokurov}, V. and {Bendjoya}, P. and {Berihuete}, A. and {Bianchi}, L. and {Bienaym{\'e}}, O. and {Billebaud}, F. and {Blagorodnova}, N. and {Blanco-Cuaresma}, S. and {Boch}, T. and {Bombrun}, A. and {Borrachero}, R. and {Bouquillon}, S. and {Bourda}, G. and {Bouy}, H. and {Bragaglia}, A. and {Breddels}, M.~A. and {Brouillet}, N. and {Br{\"u}semeister}, T. and {Bucciarelli}, B. and {Burgess}, P. and {Burgon}, R. and {Burlacu}, A. and {Busonero}, D. and {Buzzi}, R. and {Caffau}, E. and {Cambras}, J. and {Campbell}, H. and {Cancelliere}, R. and {Cantat-Gaudin}, T. and {Carlucci}, T. and {Carrasco}, J.~M. and {Castellani}, M. and {Charlot}, P. and {Charnas}, J. and {Chiavassa}, A. and {Clotet}, M. and {Cocozza}, G. and {Collins}, R.~S. and {Costigan}, G. and {Crifo}, F. and {Cross}, N.~J.~G. and {Crosta}, M. and {Crowley}, C. and {Dafonte}, C. and {Damerdji}, Y. and {Dapergolas}, A. and {David}, P. and {David}, M. and {De Cat}, P.},
        title = "{Gaia Data Release 1. Summary of the astrometric, photometric, and survey properties}",
      journal = {\aap},
         year = 2016,
        month = nov,
       volume = {595},
          eid = {A2},
        pages = {A2},
          doi = {10.1051/0004-6361/201629512},
archivePrefix = {arXiv},
       eprint = {1609.04172},
 primaryClass = {astro-ph.IM},
       adsurl = {https://ui.adsabs.harvard.edu/abs/2016A&A...595A...2G}
}

@ARTICLE{2025ApJS..279...53L,
       author = {{Li}, Xiangyu and {Chen}, Huiling and {Huang}, Yang and {Zhang}, Huawei and {Beers}, Timothy C. and {Zhu}, Linxuan and {Liu}, Jifeng},
        title = "{A Metallicity Catalog of Very Metal-poor Main-sequence Turn-off and Red Giant Stars from LAMOST DR10}",
      journal = {\apjs},
         year = 2025,
        month = aug,
       volume = {279},
       number = {2},
          eid = {53},
        pages = {53},
          doi = {10.3847/1538-4365/ade3ca},
archivePrefix = {arXiv},
       eprint = {2506.09705},
 primaryClass = {astro-ph.GA},
       adsurl = {https://ui.adsabs.harvard.edu/abs/2025ApJS..279...53L}
}

@ARTICLE{2015ApJ...799..134Y,
       author = {{Yuan}, Haibo and {Liu}, Xiaowei and {Xiang}, Maosheng and {Huang}, Yang and {Chen}, Bingqiu},
        title = "{Stellar Loci. I. Metallicity Dependence and Intrinsic Widths}",
      journal = {\apj},
         year = 2015,
        month = feb,
       volume = {799},
       number = {2},
          eid = {134},
        pages = {134},
          doi = {10.1088/0004-637X/799/2/134},
archivePrefix = {arXiv},
       eprint = {1412.1232},
 primaryClass = {astro-ph.SR},
       adsurl = {https://ui.adsabs.harvard.edu/abs/2015ApJ...799..134Y}
}

@ARTICLE{2022ApJ...925..164H,
       author = {{Huang}, Yang and {Beers}, Timothy C. and {Wolf}, Christian and {Lee}, Young Sun and {Onken}, Christopher A. and {Yuan}, Haibo and {Shank}, Derek and {Zhang}, Huawei and {Wang}, Chun and {Shi}, Jianrong and {Fan}, Zhou},
        title = "{Beyond Spectroscopy. I. Metallicities, Distances, and Age Estimates for Over 20 Million Stars from SMSS DR2 and Gaia EDR3}",
      journal = {\apj},
         year = 2022,
        month = feb,
       volume = {925},
       number = {2},
          eid = {164},
        pages = {164},
          doi = {10.3847/1538-4357/ac21cb},
archivePrefix = {arXiv},
       eprint = {2104.14154},
 primaryClass = {astro-ph.SR},
       adsurl = {https://ui.adsabs.harvard.edu/abs/2022ApJ...925..164H}
}

@ARTICLE{2023ApJ...957...65H,
       author = {{Huang}, Yang and {Beers}, Timothy C. and {Yuan}, Haibo and {Tan}, Ke-Feng and {Wang}, Wei and {Zheng}, Jie and {Li}, Chun and {Lee}, Young Sun and {Li}, Hai-Ning and {Zhao}, Jing-Kun and {Xue}, Xiang-Xiang and {Liu}, Yujuan and {Zhang}, Huawei and {Sun}, Xue-Ang and {Li}, Ji and {Gu}, Hong-Rui and {Wolf}, Christian and {Onken}, Christopher A. and {Liu}, Jifeng and {Fan}, Zhou and {Zhao}, Gang},
        title = "{Beyond Spectroscopy. II. Stellar Parameters for over 20 Million Stars in the Northern Sky from SAGES DR1 and Gaia DR3}",
      journal = {\apj},
         year = 2023,
        month = nov,
       volume = {957},
       number = {2},
          eid = {65},
        pages = {65},
          doi = {10.3847/1538-4357/ace628},
archivePrefix = {arXiv},
       eprint = {2307.04469},
 primaryClass = {astro-ph.GA},
       adsurl = {https://ui.adsabs.harvard.edu/abs/2023ApJ...957...65H}
}

@ARTICLE{2025PASA...42...51B,
       author = {{Buder}, Sven and {Kos}, Janez and {Wang}, Xi Ella and {McKenzie}, Madeleine and {Howell}, Madeleine and {Martell}, Sarah and {Hayden}, Michael R. and {Zucker}, Daniel B. and {Nordlander}, Thomas and {Montet}, Benjamin and {Traven}, Gregor and {Bland-Hawthorn}, Joss and {de Silva}, Gayandhi M. and {Freeman}, Kenneth and {Lewis}, Geraint and {Lind}, Karin and {Sharma}, Sanjib and {Simpson}, Jeffrey D. and {Stello}, Dennis and {Zwitter}, Tomaz and {Amarsi}, Anish M. and {Armstrong}, Joseph J. and {Banks}, Kirsten and {Beavis}, Mark and {Beeson}, Kevin-Luke and {Chen}, Boquan and {Ciuc{\u{a}}}, Ioana and {da Costa}, Gary S. and {de Grijs}, Richard and {Martin}, Bailey and {Nataf}, David Moise and {Ness}, Melissa and {Rains}, Adam D. and {Scarr}, Tim and {Vogrin{\v{c}}i{\v{c}}}, Rok and {Wang}, Zixian Purmortal and {Wittenmyer}, Rob A. and {Xie}, Yi Anne and {The Galah Collaboration}},
        title = "{The GALAH survey: Data release 4}",
      journal = {\pasa},
         year = 2025,
        month = may,
       volume = {42},
          eid = {e051},
        pages = {e051},
          doi = {10.1017/pasa.2025.26},
archivePrefix = {arXiv},
       eprint = {2409.19858},
 primaryClass = {astro-ph.GA},
       adsurl = {https://ui.adsabs.harvard.edu/abs/2025PASA...42...51B}
}

@software{2024zndo..13333814G,
       author = {{Gordon}, Karl},
        title = "{dust\_extinction: Interstellar Dust Extinction Models}",
         year = 2024,
        month = aug,
          eid = {10.5281/zenodo.13333814},
          doi = {10.5281/zenodo.13333814},
      version = {v1.5},
    publisher = {Zenodo},
       adsurl = {https://ui.adsabs.harvard.edu/abs/2024zndo..13333814G}
}

@ARTICLE{2023MNRAS.524.1855Z,
       author = {{Zhang}, Xiangyu and {Green}, Gregory M. and {Rix}, Hans-Walter},
        title = "{Parameters of 220 million stars from Gaia BP/RP spectra}",
      journal = {\mnras},
         year = 2023,
        month = sep,
       volume = {524},
       number = {2},
        pages = {1855-1884},
          doi = {10.1093/mnras/stad1941},
archivePrefix = {arXiv},
       eprint = {2303.03420},
 primaryClass = {astro-ph.SR},
       adsurl = {https://ui.adsabs.harvard.edu/abs/2023MNRAS.524.1855Z}
}

@ARTICLE{2023ApJS..264...14Z,
       author = {{Zhang}, Ruoyi and {Yuan}, Haibo},
        title = "{Empirical Temperature- and Extinction-dependent Extinction Coefficients for the GALEX, Pan-STARRS 1, Gaia, SDSS, 2MASS, and WISE Passbands}",
      journal = {\apjs},
         year = 2023,
        month = jan,
       volume = {264},
       number = {1},
          eid = {14},
        pages = {14},
          doi = {10.3847/1538-4365/ac9dfa},
archivePrefix = {arXiv},
       eprint = {2210.15918},
 primaryClass = {astro-ph.GA},
       adsurl = {https://ui.adsabs.harvard.edu/abs/2023ApJS..264...14Z}
}

@ARTICLE{2025ApJS..280...15W,
       author = {{Wang}, Tao and {Yuan}, Haibo and {Chen}, Bingqiu and {Xiang}, Maosheng and {Zhang}, Ruoyi and {Huang}, Bowen and {Gu}, Hongrui and {Wang}, Shuaicong and {Li}, Jiawei},
        title = "{An All-sky 3D Dust Map Based on Gaia and LAMOST}",
      journal = {\apjs},
         year = 2025,
        month = sep,
       volume = {280},
       number = {1},
          eid = {15},
        pages = {15},
          doi = {10.3847/1538-4365/adea39},
archivePrefix = {arXiv},
       eprint = {2509.07640},
 primaryClass = {astro-ph.GA},
       adsurl = {https://ui.adsabs.harvard.edu/abs/2025ApJS..280...15W}
}

@ARTICLE{2025ApJS..279....7Y,
       author = {{Yang}, Lin and {Yuan}, Haibo and {Huang}, Bowen and {Zhang}, Ruoyi and {Beers}, Timothy C. and {Xiao}, Kai and {Xu}, Shuai and {Huang}, Yang and {Xiang}, Maosheng and {Zhang}, Meng and {Zhang}, Jinming},
        title = "{Metallicities of 20 Million Giant Stars Based on Gaia XP Spectra}",
      journal = {\apjs},
         year = 2025,
        month = jul,
       volume = {279},
       number = {1},
          eid = {7},
        pages = {7},
          doi = {10.3847/1538-4365/add5e3},
archivePrefix = {arXiv},
       eprint = {2505.05281},
 primaryClass = {astro-ph.SR},
       adsurl = {https://ui.adsabs.harvard.edu/abs/2025ApJS..279....7Y}
}

@ARTICLE{2023ApJS..267....8A,
       author = {{Andrae}, Ren{\'e} and {Rix}, Hans-Walter and {Chandra}, Vedant},
        title = "{Robust Data-driven Metallicities for 175 Million Stars from Gaia XP Spectra}",
      journal = {\apjs},
         year = 2023,
        month = jul,
       volume = {267},
       number = {1},
          eid = {8},
        pages = {8},
          doi = {10.3847/1538-4365/acd53e},
archivePrefix = {arXiv},
       eprint = {2302.02611},
 primaryClass = {astro-ph.SR},
       adsurl = {https://ui.adsabs.harvard.edu/abs/2023ApJS..267....8A}
}

@ARTICLE{2024A&A...692A.115M,
       author = {{Martin}, Nicolas F. and {Starkenburg}, Else and {Yuan}, Zhen and {Fouesneau}, Morgan and {Ardern-Arentsen}, Anke and {De Angeli}, Francesca and {Gran}, Felipe and {Montelius}, Martin and {Rusterucci}, Samuel and {Andrae}, Ren{\'e} and {Bellazzini}, Michele and {Montegriffo}, Paolo and {Esselink}, Anna F. and {Zhang}, Hanyuan and {Venn}, Kim A. and {Viswanathan}, Akshara and {Aguado}, David S. and {Battaglia}, Giuseppina and {Bayer}, Manuel and {Bonifacio}, Piercarlo and {Caffau}, Elisabetta and {C{\^o}t{\'e}}, Patrick and {Carlberg}, Raymond and {Fabbro}, S{\'e}bastien and {Fern{\'a}ndez-Alvar}, Emma and {Gonz{\'a}lez Hern{\'a}ndez}, Jonay I. and {Gonz{\'a}lez Rivera de La Vernhe}, Isaure and {Hill}, Vanessa and {Ibata}, Rodrigo A. and {Jablonka}, Pascale and {Kordopatis}, Georges and {Lardo}, Carmela and {McConnachie}, Alan W. and {Navarrete}, Camila and {Navarro}, Julio and {Recio-Blanco}, Alejandra and {S{\'a}nchez-Janssen}, Rub{\'e}n and {Sestito}, Federico and {Thomas}, Guillaume F. and {Vitali}, Sara and {Youakim}, Kristopher},
        title = "{The Pristine survey: XXIII. Data Release 1 and an all-sky metallicity catalogue based on Gaia DR3 BP/RP spectro-photometry}",
      journal = {\aap},
         year = 2024,
        month = dec,
       volume = {692},
          eid = {A115},
        pages = {A115},
          doi = {10.1051/0004-6361/202347633},
archivePrefix = {arXiv},
       eprint = {2308.01344},
 primaryClass = {astro-ph.GA},
       adsurl = {https://ui.adsabs.harvard.edu/abs/2024A&A...692A.115M}
}

@ARTICLE{2025ApJ...989L..18L,
       author = {{Limberg}, Guilherme and {Placco}, Vinicius M. and {Ji}, Alexander P. and {Yao}, Yupeng and {Chiti}, Anirudh and {Mardini}, Mohammad K. and {Frebel}, Anna and {Rossi}, Silvia},
        title = "{Discovery of an [Fe/H] {\ensuremath{\sim}} {\ensuremath{-}}4.8 Star in Gaia XP Spectra}",
      journal = {\apjl},
         year = 2025,
        month = aug,
       volume = {989},
       number = {1},
          eid = {L18},
        pages = {L18},
          doi = {10.3847/2041-8213/adf196},
archivePrefix = {arXiv},
       eprint = {2508.00067},
 primaryClass = {astro-ph.GA},
       adsurl = {https://ui.adsabs.harvard.edu/abs/2025ApJ...989L..18L}
}

@ARTICLE{2021A&A...652A..86C,
       author = {{Carrasco}, J.~M. and {Weiler}, M. and {Jordi}, C. and {Fabricius}, C. and {De Angeli}, F. and {Evans}, D.~W. and {van Leeuwen}, F. and {Riello}, M. and {Montegriffo}, P.},
        title = "{Internal calibration of Gaia BP/RP low-resolution spectra}",
      journal = {\aap},
         year = 2021,
        month = aug,
       volume = {652},
          eid = {A86},
        pages = {A86},
          doi = {10.1051/0004-6361/202141249},
archivePrefix = {arXiv},
       eprint = {2106.01752},
 primaryClass = {astro-ph.IM},
       adsurl = {https://ui.adsabs.harvard.edu/abs/2021A&A...652A..86C}
}

@ARTICLE{2024ApJS..271...13H,
       author = {{Huang}, Bowen and {Yuan}, Haibo and {Xiang}, Maosheng and {Huang}, Yang and {Xiao}, Kai and {Xu}, Shuai and {Zhang}, Ruoyi and {Yang}, Lin and {Niu}, Zexi and {Gu}, Hongrui},
        title = "{A Comprehensive Correction of the Gaia DR3 XP Spectra}",
      journal = {\apjs},
         year = 2024,
        month = mar,
       volume = {271},
       number = {1},
          eid = {13},
        pages = {13},
          doi = {10.3847/1538-4365/ad18b1},
archivePrefix = {arXiv},
       eprint = {2401.12006},
 primaryClass = {astro-ph.IM},
       adsurl = {https://ui.adsabs.harvard.edu/abs/2024ApJS..271...13H}
}

@ARTICLE{2021ApJ...912L..32P,
       author = {{Placco}, Vinicius M. and {Roederer}, Ian U. and {Lee}, Young Sun and {Almeida-Fernandes}, Felipe and {Herpich}, F{\'a}bio R. and {Perottoni}, H{\'e}lio D. and {Schoenell}, William and {Ribeiro}, Tiago and {Kanaan}, Antonio},
        title = "{SPLUS J210428.01-004934.2: An Ultra Metal-poor Star Identified from Narrowband Photometry}",
      journal = {\apjl},
         year = 2021,
        month = may,
       volume = {912},
       number = {2},
          eid = {L32},
        pages = {L32},
          doi = {10.3847/2041-8213/abf93d},
archivePrefix = {arXiv},
       eprint = {2105.04573},
 primaryClass = {astro-ph.SR},
       adsurl = {https://ui.adsabs.harvard.edu/abs/2021ApJ...912L..32P}
}

@ARTICLE{2025arXiv250921643J,
       author = {{Ji}, Alexander P. and {Chandra}, Vedant and {Mejias-Torres}, Selenna and {Zhang}, Zhongyuan and {Eitner}, Philipp and {Schlaufman}, Kevin C. and {Andales}, Hillary Diane and {Do}, Ha and {Orrantia}, Natalie M. and {Tudmilla}, Rithika and {Thibodeaux}, Pierre N. and {Stassun}, Keivan G. and {Howell}, Madeline and {Tayar}, Jamie and {Bergemann}, Maria and {Casey}, Andrew R. and {Johnson}, Jennifer A. and {Carlberg}, Joleen K. and {Cerny}, William and {Fernandez-Trincado}, Jose G. and {Hawkins}, Keith and {Kollmeier}, Juna A. and {Laporte}, Chervin F.~P. and {Limberg}, Guilherme and {Matsuno}, Tadafumi and {Meszaros}, Szabolcs and {Morrison}, Sean and {Nidever}, David L. and {Stringfellow}, Guy S. and {Schneider}, Donald P. and {Thai}, Riley},
        title = "{A nearly pristine star from the Large Magellanic Cloud}",
      journal = {arXiv e-prints},
         year = 2025,
        month = sep,
          eid = {arXiv:2509.21643},
        pages = {arXiv:2509.21643},
          doi = {10.48550/arXiv.2509.21643},
archivePrefix = {arXiv},
       eprint = {2509.21643},
 primaryClass = {astro-ph.SR},
       adsurl = {https://ui.adsabs.harvard.edu/abs/2025arXiv250921643J}
}

@software{daniela_ruz_mieres_2022_7015044,
  author       = {Daniela Ruz-Mieres},
  title        = {gaia-dpci/GaiaXPy: GaiaXPy 1.2.0},
  month        = aug,
  year         = 2022,
  publisher    = {Zenodo},
  version      = {1.2.0},
  doi          = {10.5281/zenodo.7015044},
  url          = {https://doi.org/10.5281/zenodo.7015044},
}

@ARTICLE{2024A&A...683L..11V,
       author = {{Viswanathan}, Akshara and {Starkenburg}, Else and {Matsuno}, Tadafumi and {Venn}, Kim A. and {Martin}, Nicolas F. and {Longeard}, Nicolas and {Ardern-Arentsen}, Anke and {Carlberg}, Raymond G. and {Fabbro}, S{\'e}bastien and {Kordopatis}, Georges and {Montelius}, Martin and {Sestito}, Federico and {Yuan}, Zhen},
        title = "{Gaia's brightest very metal-poor (VMP) stars. Metallicity catalogue of a thousand VMP stars from Gaia's radial velocity spectrometer spectra}",
      journal = {\aap},
         year = 2024,
        month = mar,
       volume = {683},
          eid = {L11},
        pages = {L11},
          doi = {10.1051/0004-6361/202347944},
archivePrefix = {arXiv},
       eprint = {2309.06137},
 primaryClass = {astro-ph.GA},
       adsurl = {https://ui.adsabs.harvard.edu/abs/2024A&A...683L..11V}
}

@ARTICLE{2023A&A...674A.194B,
       author = {{Bellazzini}, M. and {Massari}, D. and {De Angeli}, F. and {Mucciarelli}, A. and {Bragaglia}, A. and {Riello}, M. and {Montegriffo}, P.},
        title = "{Photometric metallicity for 694 233 Galactic giant stars from Gaia DR3 synthetic Str{\"o}mgren photometry. Metallicity distribution functions of halo substructures}",
      journal = {\aap},
         year = 2023,
        month = jun,
       volume = {674},
          eid = {A194},
        pages = {A194},
          doi = {10.1051/0004-6361/202345921},
archivePrefix = {arXiv},
       eprint = {2304.10772},
 primaryClass = {astro-ph.GA},
       adsurl = {https://ui.adsabs.harvard.edu/abs/2023A&A...674A.194B}
}

@ARTICLE{2025A&A...700A..74O,
       author = {{Omkumar}, Abinaya O. and {Cioni}, Maria-Rosa L. and {Subramanian}, Smitha and {de Bruijne}, Jos and {Nair}, Aparna and {Dias}, Bruno},
        title = "{Str{\"o}mgren photometric metallicity map of the Magellanic Cloud stars using Gaia DR3─XP spectra}",
      journal = {\aap},
         year = 2025,
        month = aug,
       volume = {700},
          eid = {A74},
        pages = {A74},
          doi = {10.1051/0004-6361/202452510},
archivePrefix = {arXiv},
       eprint = {2506.10749},
 primaryClass = {astro-ph.GA},
       adsurl = {https://ui.adsabs.harvard.edu/abs/2025A&A...700A..74O}
}

@ARTICLE{2024ApJ...963...95C,
       author = {{Cunningham}, Emily C. and {Hunt}, Jason A.~S. and {Price-Whelan}, Adrian M. and {Johnston}, Kathryn V. and {Ness}, Melissa K. and {Lu}, Yuxi (Lucy) and {Escala}, Ivanna and {Stelea}, Ioana A.},
        title = "{Chemical Cartography of the Sagittarius Stream with Gaia}",
      journal = {\apj},
         year = 2024,
        month = mar,
       volume = {963},
       number = {2},
          eid = {95},
        pages = {95},
          doi = {10.3847/1538-4357/ad187b},
archivePrefix = {arXiv},
       eprint = {2307.08730},
 primaryClass = {astro-ph.GA},
       adsurl = {https://ui.adsabs.harvard.edu/abs/2024ApJ...963...95C}
}

@ARTICLE{2021AJ....161..147B,
       author = {{Bailer-Jones}, C.~A.~L. and {Rybizki}, J. and {Fouesneau}, M. and {Demleitner}, M. and {Andrae}, R.},
        title = "{Estimating Distances from Parallaxes. V. Geometric and Photogeometric Distances to 1.47 Billion Stars in Gaia Early Data Release 3}",
      journal = {\aj},
         year = 2021,
        month = mar,
       volume = {161},
       number = {3},
          eid = {147},
        pages = {147},
          doi = {10.3847/1538-3881/abd806},
archivePrefix = {arXiv},
       eprint = {2012.05220},
 primaryClass = {astro-ph.SR},
       adsurl = {https://ui.adsabs.harvard.edu/abs/2021AJ....161..147B}
}
\bibliographystyle{aasjournal}

%% This command is needed to show the entire author+affiliation list when
%% the collaboration and author truncation commands are used.  It has to
%% go at the end of the manuscript.
%\allauthors

%% Include this line if you are using the \added, \replaced, \deleted
%% commands to see a summary list of all changes at the end of the article.
%\listofchanges

\appendix

\section{Calibration} \label{sec:appendix_calibration}
The GALAH DR4 sample is cross-matched with the SAGA/PASTEL HRS sample, and the common giant and dwarf stars are analyzed separately to examine the metallicity scale of SAGA/PASTEL HRS. The results are shown in Figure~\ref{fig_app_result}. Overall, the metallicities derived from the HRS sample for both giant and dwarf stars are broadly consistent with those from GALAH DR4, but show a systematic offset toward lower metallicities. Two-order polynomial functions are adopted to characterize this systematic offset.

\setcounter{figure}{0}            
\renewcommand{\thefigure}{A\arabic{figure}}

\begin{figure}[htbp]
\centering
\includegraphics[width=0.7\textwidth]{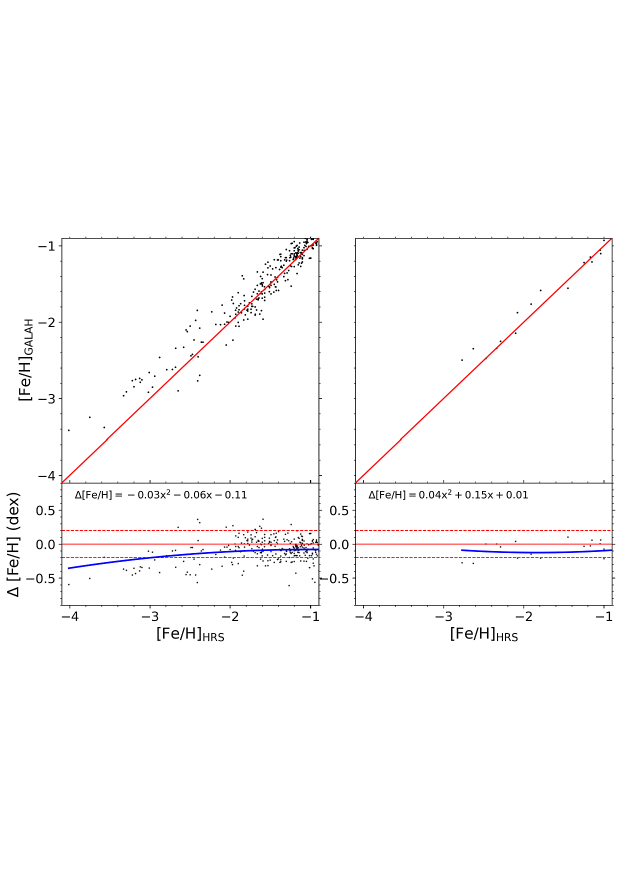} % 占一半
%\caption{Comparison of the metallicity for common giant (left panel) %and dwarf (right panel) stars between SAGA/PASTEL high resolution spectra and GALAH DR4. The lower part of each panel shows the metallicity difference (metallicity from HRS spectra minus metallicity from GALAH DR4) as a function of HRS metallicity. The blue line (with the function marked in the top corner, here $x$ denotes [Fe/H] from HRS spectra) shows the two order polynomial fits to the data points.
\label{fig_app_result}
\end{figure}

\end{document}